\documentstyle[stwol]{article}

\input{psfig}

\def\jour#1#2#3#4{{#1} {\bf #2}, #3 (#4)}

\def\nima{{\em Nucl. Instrum. Methods} A}
\def\npb{{\em Nucl. Phys.} B}
\def\plb{{\em Phys. Lett.}  B}
\def\prl{\em Phys. Rev. Lett.}
\def\prd{{\em Phys. Rev.} D}
\def\zpc{{\em Z. Phys.} C}
\def\ptp{{\em Prog. Theor. Phys.}}
\def\rmp{{ \em Rev. Mod. Phys.}}

\def\vep{\varepsilon}

\def\be{\begin{equation}}
\def\ee{\end{equation}}
\def\bea{\begin{eqnarray}}
\def\eea{\end{eqnarray}}

\def\ho{H^0}
\def\hb{\bar{H}^0}

\def\pio{\pi^0}
\def\pip{\pi^+}

\def\bo{B^0}
\def\bb{\bar{B}^0}
\def\bp{B^+}
\def\bm{B^-}
\def\bos{B^0_s}
\def\bbs{\bar{B}^0_s}
\def\bod{B^0_d}
\def\bbd{\bar{B}^0_d}
\def\bs{B_s}
\def\bd{B_d}
\def\fbs{f_{\bs}}

\def\boket{\left|\bo\right>}
\def\bbket{\left|\bb\right>}
\def\bsket{\left|B_L\right>}
\def\blket{\left|B_S\right>}
\def\botim{\left|\bo(t)\right>}
\def\bbtim{\left|\bb(t)\right>}

\def\kstgam{K^*\gamma}
\def\rhogam{\rho\gamma}
\def\rhomgam{\rho^-\gamma}

\def\romgam{\rho^0/\omega\gamma}

\def\pln{\pi\ell\nu}
\def\rwln{\rho/\omega\ell\nu}
\def\rln{\rho\ell\nu}
\def\emiss{E_{\rm\scriptsize miss}}
\def\pmiss{\vec{P}_{\rm\scriptsize miss}}
\def\mmiss{M^2_{\rm\scriptsize miss}}
\def\mcand{M_{m\ell\nu}}
\def\ebeam{E_{\rm\scriptsize beam}}
\def\ecand{E_{m\ell\nu}}
\def\dele{\Delta E}

\def\ds{D^*}

\def\kp{K^+}

\def\pnn{\pi^+\nu\bar{\nu}}

\def\zo{Z^0}

\def\vub{V_{ub}}
\def\vus{V_{us}}
\def\vud{V_{ud}}
\def\vtb{V_{tb}}
\def\vts{V_{ts}}
\def\vtd{V_{td}}
\def\vcb{V_{cb}}
\def\vcs{V_{cs}}
\def\vcd{V_{cd}}

\def\delm{\Delta m}
\def\delmd{\delm_d}
\def\delms{\delm_s}

\def\gamd{\Gamma_d}

\def\qj{Q_J}
\def\qjo{\qj^{\rm\scriptsize opp}}
\def\qjs{\qj^{\rm\scriptsize same}}

\def\rhohat{\hat{\rho}}
\def\chat{\hat{c}}

\def\nubar{\bar{\nu}}
\def\etal{{\it et al.}}
\def\ten#1{10^{#1}}
\def\tten#1{\times 10^{#1}}
\def\calb{{\cal B}}
\def\prob{{\cal P}}
\def\like{{\cal L}}
\def\form{{\cal F}}
\def\orde{{\cal O}}
\def\mevc{MeV/$c$}

\def\deg{^\circ}

\begin{document}

\begin{titlepage}
{\hfill UR-1494}\\
{\hspace*{\fill} {April, 1997}}\\
{\hspace*{\fill} {hep-ex/9704017}}\\
\vspace{2.5 cm}

\begin{center}
{\bf STATUS OF WEAK QUARK MIXING$^\dag$}\\
\vspace{5 em}

{\bf Lawrence K. Gibbons }\\
\vspace{0.75 em}

{\it Department of Physics and Astronomy, University of Rochester}\\
{\it Rochester, NY 14627, USA}\\
\vspace{8 em}

{\bf Abstract}\\
\end{center}
{The experimental status of the Cabibbo-Kobayashi-Maskawa matrix is reviewed. 
Measurements discussed include
$B^0_{(s)}-\bar{B}^0_{(s)}$ mixing and several rare $B$ and
$K$ decays with implications for $|V_{td}|$ 
and $|V_{ts}|$.  Extraction of $|\vcb|$ and $|\vub|$ from studies of
semileptonic $B$ decay is also discussed.}\\
\vspace{20 em}

{$^\dag$\footnotesize Invited talk given at the 28th
International Conference on High Energy Physics, Warsaw, Poland, 25-31 July
1996, to appear in the proceedings.}
\setcounter{footnote}{0}
\end{titlepage}

\title{STATUS OF WEAK QUARK MIXING}

\author{ L.K. GIBBONS }

\address{University of Rochester, Rochester, NY 14627, USA}


\twocolumn[\maketitle\abstracts{
{The experimental status of the Cabibbo-Kobayashi-Maskawa matrix is reviewed. 
Measurements discussed include
$B^0_{(s)}-\bar{B}^0_{(s)}$ mixing and several rare $B$ and
$K$ decays with implications for $|V_{td}|$ 
and $|V_{ts}|$.  Extraction of $|\vcb|$ and $|\vub|$ from studies of
semileptonic $B$ decay is also discussed.}}]

\section{Introduction}
This review summarizes the status of weak quark mixing, focusing on
our knowledge of the Cabibbo-Kobayashi-Maskawa (CKM) 
matrix~\cite{bb:ckm_a,bb:ckm_b}
elements $V_{cb}$, $V_{ub}$, $V_{td}$ and $V_{ts}$.  Precise evaluation of
these elements is crucial to our understanding of the origins of $CP$
violation --- in particular, whether the $CP$--violating phase of the CKM 
matrix is sufficient to explain the observed rate of the $CP$-violating 
decay $K_L\to\pi\pi$.

The results discussed
here are intimately connected with topics covered in more detail in three
other talks.  To extract the CKM elements, we must understand the dynamics
underlying the decays studied; J.~Richman~\cite{bb:jeffs-talk} and
G.~Martinelli~\cite{bb:guidos-talk} discuss the dynamics in detail.  The
elements under discussion are the basic inputs to the ``Unitarity Triangle''
(UT) analysis to test the hypothesis that the rate of $CP$ violation observed
in the neutral kaon system is consistent with arising solely from a
$CP$--violating phase in the CKM matrix. A.~Buras~\cite{bb:andrzejs-talk}
will present a detailed UT analysis in his talk.

The CKM matrix appears in the weak charged current, 
\be
J_\mu = (\bar{u} \bar{c} \bar{t})_L\gamma_\mu 
\left(\!\!\begin{array}{rrr} \vud & \vus & \vub \\ 
                             \vcd & \vcs & \vcb \\ 
                             \vtd & \vts & \vtb
\end{array}\!\!\right)
\left(\!\!\begin{array}{c} d \\ s \\ b \end{array}\!\!\right)_L,
\ee
rotating the quark system from the flavor eigenstate basis to the weak 
eigenstate  basis.  A convenient parameterization, due to
Wolfenstein~\cite{bb:wolfen}, in which the matrix is expanded in powers
of the Cabibbo angle $\lambda=|\vtd|=0.22$ is
\bea
\lefteqn{V_{\scriptscriptstyle CKM} \approx} \\
 & & \!\!\!\left(\!\!\begin{array}{ccc}
1-\lambda^2/2 & \lambda & A\lambda^3(\rho-i\eta) \\
-\lambda & 1-\lambda^2/2 & A\lambda^2 \\
 A\lambda^3(1-\rho-i\eta)  & -A\lambda^2 & 1
 \end{array}\!\!\!\right).  \nonumber 
\eea
Within this approximation, good to order $\orde(\lambda^3)$,
$CP$--violating amplitudes are proportional to the parameter $\eta$.  
Buras~\cite{bb:andrzejs-talk} discusses both this parameterization and the
unitary triangle that results from the unitary property $\sum V_{ik}V^*_{jk}=0$
for $i\neq j$ in some detail.

Section 2 of this review will discuss experimental measurements sensitive to
$|V_{td}|$ and $|V_{ts}|$.  Section 3 will summarize the determination of
$|\vcb|$ from inclusive and exclusive $b\to c\ell\nu$ decays, while section 4
will focus on the new determination of $|\vub|$ from exclusive $B\to X_u\ell\nu$
studies.

\section{$V_{td}$ and $V_{ts}$}
The top quark presents an experimental irony.  Its large mass and short
lifetime make possible an accurate determination of its mass with only
a handful of events.  In fact, its mass is fractionally among the best 
determined for any quark.  However, the large mass 
limits us to a total event sample of only a few events, making direct 
determination of $|V_{td}|$ and $|V_{ts}|$ impossible at this time.  
We must therefore resort to
indirect means of determining those CKM elements.

\begin{figure}[tb]
\centering
\psfig{figure=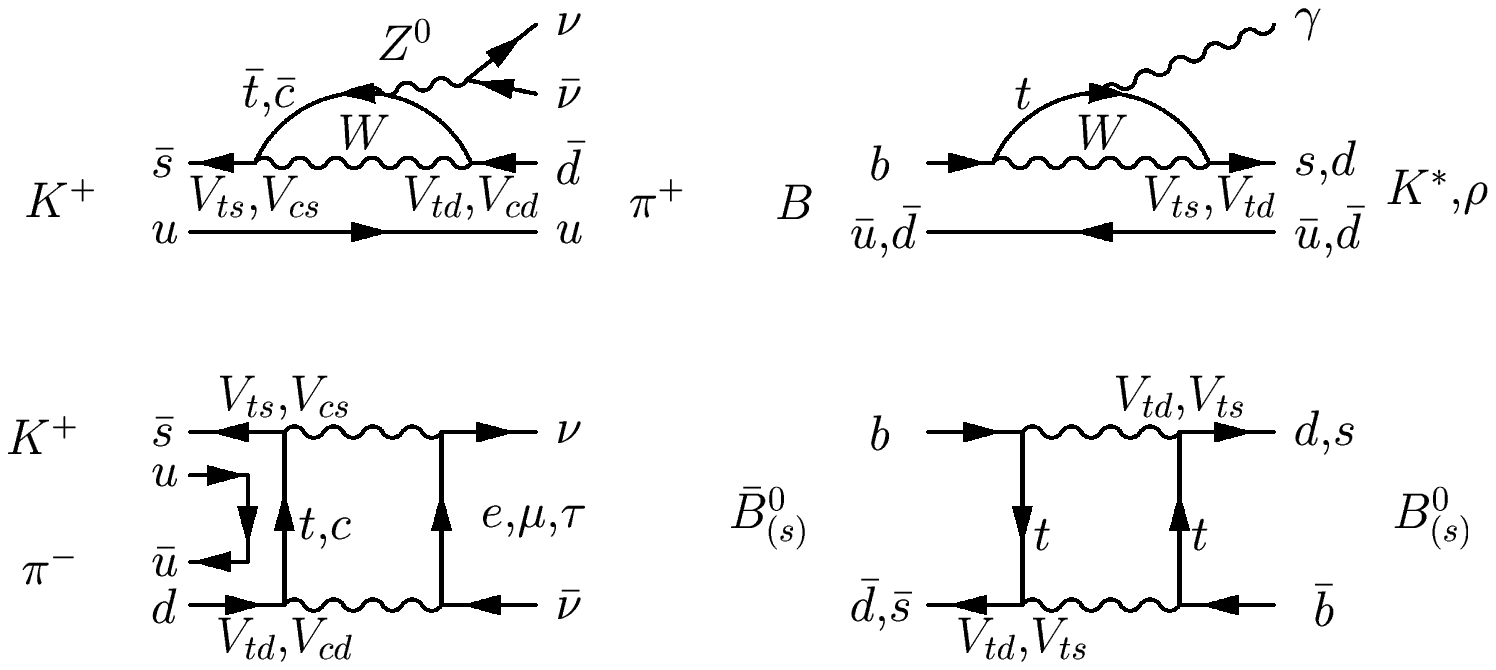,width=7.5cm}
\caption{Diagrams for processes from which information about $\vtd$ and 
$\vts$ can be inferred: $\kp\to\pnn$ (left), 
$B\to K^*(\rho)\gamma$ (top right), and $\bo-\bb$ mixing (bottom right).}
\label{fig:top-loops}
\end{figure}

Weak processes that contain a virtual top quark in a loop provide such a
means. Because the amplitudes are roughly proportional to the square of
the large top
quark mass, such processes can have accessible rates.  In fact, the large
rate observed in the initial $\bo-\bb$ mixing
measurements~\cite{bb:first-argus-mix} was the first hint
that the top quark was unusually heavy.  Figure~\ref{fig:top-loops} shows a
variety of these processes.  Electroweak penguin diagrams contribute to the
decays $B\to \kstgam$, $B\to \rhogam$ and $\kp\to\pnn$.  The box diagram, a
second order weak process, also contributes to the decay $\kp\to\pnn$, as
well as to $\bo-\bb$ and $\bos-\bbs$ mixing.  The experimental status
for these four different processes follows, with a summary of the implications
for $|V_{td}|$ and $|V_{ts}|$ at the end of this section.

\subsection{$K^+\to\pi^+\nu\bar{\nu}$}
Theoretically, one of the cleanest avenues for the extraction of $|\vts|$ is
the semileptonic decay $K^+\to\pi^+\nu\bar{\nu}$.  Long distance corrections
have been found to be negligible~\cite{bb:knunu_a,bb:knunu_b} compared to the
short distance contribution from diagrams like those in
Figure~\ref{fig:top-loops}. Uncertainties in the hadronic current can be
eliminated by normalizing to the related semileptonic decay $\kp\to\pio
e^+\nu$.  For this decay, while the amplitudes of diagrams containing the top
quark are enhanced by its large mass, they also contain the factor
$\vts^*\vtd$, of order $\lambda^5$.  On the other hand, the amplitudes from
analogous diagrams with a charm quark in the loop contain the factor
$\vcs^*\vcd$, of order $\lambda$, making the charmed loops competitive
with the top.  The charm quark contribution still accounts for the largest
theoretical uncertainty, though that uncertainty has recently been greatly 
reduced to about 4\% by a next-to-leading order calculation of Buchalla and
Buras~\cite{bb:knunu_c}.  Within the standard model,
predictions~\cite{bb:andrzejs-talk} for this decay rate lie in the range
$(0.6-1.2)\tten{-10}$.

\begin{figure}[tb]
\centering
\psfig{figure=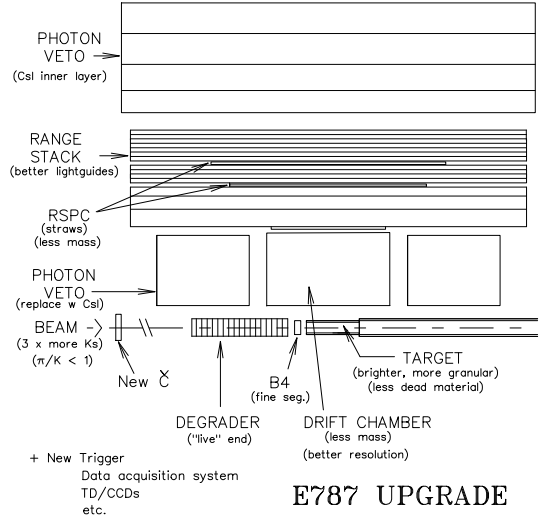,width=7.0cm}
\caption{Schematic of the upgraded E787 detector.}
\label{fig:e787}
\end{figure}

\begin{figure}[tb]
\centering
\psfig{figure=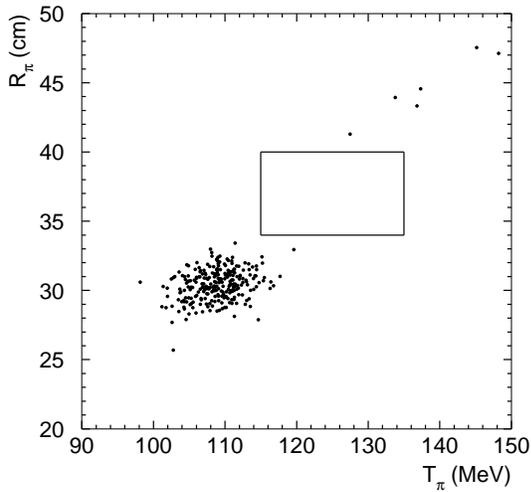,width=7.0cm}
\caption{Range ($R_\pi$) versus kinetic energy ($T_\pi$) for events 
satisfying the
E787 selection criteria and $211\le p_\pi \le 243$ \mevc.  The box
indicates the $\kp\to\pip\nu\nubar$ search region. $\kp\to\pip\pio$ 
background populates the lower left region;
$\kp\to\mu^+\nu$ populates the upper right.}
\label{fig:e787_data}
\end{figure}

The Brookhaven experiment E787 has been searching for this decay with an
elegant detector, shown in its current upgraded form in Figure~\ref{fig:e787}.
A high intensity beam from the AGS delivers $\kp$ mesons, which are tagged in
a \v{C}erenkov counter, slowed and then stopped in an active segmented
target.  Charged particles from the target are tracked in a drift chamber and
range out in a segmented range stack of plastic scintillator layers.  Having
instrumented all active channels with transient digitizers, E787 can
suppress background from $\kp\to\mu^+\nu$ decays (63\% branching fraction) by
identifying the complete decay sequence $\pip\to\mu^+\to e^+$.  
Photon vetos suppress
$\pip\pio$ decays (21\% branching fraction).  Candidate $\pip$ are selected
in a momentum range between the two peaks from $\mu^+\nu$ and $\pip\pio$
decays.  The final data distribution~\cite{bb:knunu_d} in range ($R$) versus
kinetic energy ($T$) measured in the range stack is shown in
Figure~\ref{fig:e787_data} for the 1989-1991 runs.  No events are observed in
the signal region.  From this data E787 has set the 90\% C.L. upper limit
$\calb(\kp\to\pip\nu\nubar)<2.4\tten{-9}$.

The E787 experiment has been accumulating more data with its upgraded detector.  
The upgrade has resulted in a detector with lower mass and better light
collection, and hence
with better $\pi/\mu$ separation in both $T$ and $R$ and with improved
photon vetoing.  After initial runs in 1995 and 1996, E787 estimates a
factor of six improvement in reach over their previous data set.  After
completing their planned 1997 run, they expect a final sensitivity of
about $2\tten{-10}$. The theoretical uncertainties in this mode are small
enough that a signal at this level would be a strong indication of physics
beyond the standard model.

\subsection{$B\to K^*\gamma$, $B\to\rho\gamma$}

Study of the exclusive radiative penguin decays $B\to\kstgam$ and
$B\to\rhogam$ can play several roles in constraining CKM elements.
Foremost is the extraction of the ratio $|\vtd/\vts|$ from the ratio of
branching fractions $R\equiv\calb(B\to\rhogam)/\calb(B\to\kstgam)$, though
one must correct for differences in phase space, $SU(3)$-breaking, and,
particularly for $\rhogam$, long distance (LD) contributions

Recent estimates have indicated that the long distance contributions may be
manageable: of order~\cite{bb:kst_a,bb:kst_b} 10\% for $\bm\to\rhomgam$,
and a few percent~\cite{bb:kst_c} or less for $\bo\to\romgam$.  The
question, however, is still under active investitarion. The
phase space correction~\cite{bb:kst_d} is small, $1.02\pm0.02$.  The
$SU(3)$-breaking correction $\xi$ is model-dependent, but this dependence
should be smaller than for predictions of the individual $\rhogam$ and
$\kstgam$ rates.  
Calculations~\cite{bb:kst_d,bb:kst_e,bb:kst_f} of $\xi$ range from 0.58 to
0.90.

\begin{figure}[tb]
\centering
\psfig{figure=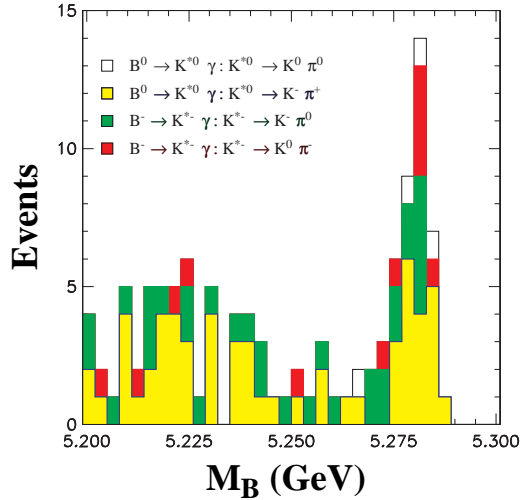,width=7.0cm}
\caption{Reconstructed mass distribution of $B\to\kstgam$ candidates in
the four modes studied at CLEO.}
\label{fig:kstgam}
\end{figure}

CLEO~\cite{bb:kst-cleo} has recently updated its measurements of $B\to\kstgam$
and placed new upper limits on $\bm\to\rhomgam$ and $\bo\to\romgam$.  A strong
signal can be seen in the sum of the four $\kstgam$ modes examined
(Figure~\ref{fig:kstgam}), and combined they yield
$\calb(B\to\kstgam)=(4.2\pm0.8\pm0.6)\tten{-5}$.  CLEO sees no evidence for a
signal in any of the three $b\to d\gamma$ modes studied.  Combining the
$\kstgam$ measurement with these
three modes under the assumption that
$\Gamma(\bm\to\rhomgam)=2\Gamma(\bo\to\romgam)$,
CLEO obtains the 90\% C.L. upper limit $R<0.19$.  Assuming no long distance
contribution to any of the $b\to d\gamma$ modes, but correcting for phase
space and $SU(3)$--breaking, this limit implies $|\vtd/\vts|<0.45-0.56$, where
the range comes from the different predictions for $\xi$.  Allowing for a
10\% long distance amplitude for $\bm\to\rhomgam$ that, to be conservative,
interferes destructively, the limit's upper range would
change to $|\vtd/\vts|<0.59$.

\subsection{$\bo-\bb$ mixing}

The formalism for mixing of neutral $B$ mesons, either $\bs$ or $\bd$, is
completely analogous to that of neutral $K$ mesons.  The weak eigenstates are
a mixture of the flavor $\boket$ and $\bbket$ eigenstates: 
\bea
\blket & = & p\boket + q\bbket, \nonumber \\ 
\bsket & = & p\boket - q\bbket,
\eea 
with the complex amplitudes $p$ and $q$ normalized such that $|p|^2+|q|^2=1$.
The labeling of the weak neutral $B$ eigenstates analogously to their
kaon counterparts is somewhat of a misnomer.  For a neutral meson $H$,
the only contribution to the lifetime difference of the weak eigenstates comes
from final states that are common to both $\ho$ and $\hb$.  In the case of the
kaon, the $\pi\pi$ final states are by far the dominant final state, and their
contribution leads to a $K_L$ lifetime about 580 times larger than the $K_S$
lifetime.  In the case of the $B$ mesons~\cite{bb:mix-bigi}, the branching
fractions of final states common to both the $\bo$ and $\bb$ are only of order
$\ten{-3}$, and different final states will contribute to the
width with different signs.  Hence the lifetime difference
$\Delta\Gamma=\Gamma_L-\Gamma_S$ is expected to be quite small, with
$|\Delta\Gamma|/\Gamma$ of order a few percent or less.

In the $\bod-\bbd$ system, where the branching fractions for
a large number of decay modes have been measured, the common final states do
have small branching fractions.  This is expected to be
the case in the $\bos-\bbs$ system as well.  I will assume that
$\Delta\Gamma=0$, as the experiments do, which implies~\cite{bb:mix-bigi}
that the ratio $|p/q|=1$.

Processes such as the box diagrams in Figure~\ref{fig:top-loops}
do generate a mass difference, 
$\delm\equiv M_{B_L}-M_{B_S}$, just as in the neutral kaon system.  Given this
mass difference, an initially pure $\bo$ or $\bb$ state evolves as a function
of proper time according to
\bea
\lefteqn{\botim = e^{-t(\frac{\Gamma}{2}+iM)}\times} \nonumber \\
 & &   \left[\cos(\frac{\delm t}{2})\boket + 
 \frac{iq}{p}\sin(\frac{\delm t}{2})\bbket\right] \nonumber \\
\lefteqn{\bbtim = e^{-t(\frac{\Gamma}{2}+iM)}\times}  \\
 & &   \left[\cos(\frac{\delm t}{2})\bbket + 
 \frac{ip}{q}\sin(\frac{\delm t}{2})\boket\right], \nonumber
\eea
where $M\equiv(M_L+M_S)/2$.  The probability as a function of proper time
that an initially pure $\bo$ or $\bb$ state decay as the opposite, or 
``mixed'', state is
\be
\prob_m(t) = \frac{\Gamma}{2}e^{-\Gamma t}[1-\cos(\delm t)]
\label{eq:mixprob}
\ee
and as the same, unmixed state is
\be
\prob_u(t) = \frac{\Gamma}{2}e^{-\Gamma t}[1+\cos(\delm t)].
\ee
These forms result when $\Delta\Gamma=0$, and hence $|p/q|=1$, without assuming
strict $CP$-conservation.  Integrated over time, the total fraction $\chi_q$ of
mixed decays from an initially pure $B_q$ state is given by
\be
\chi_q=\int_0^\infty \prob_m(t)\,dt = \frac{x_q^2}{2(1+x_q^2)},
\ee
where $x_q=\delm_q/\Gamma_q$.

The ultimate experimental goal is a precise determination of the mass
differences $\delmd$ and $\delms$, from which we can obtain information
concerning $\vtd$ and $\vts$.  Within the Standard Model, the box
diagrams with a top quark in the internal loop dominate the contribution
to the mass difference.  Their evaluation for either the $B_d$ 
($q=d$) or $B_s$ ($q=s$) system yields~\cite{bb:mix-buras-1}
\be
\delm_q=\frac{G_F^2}{6\pi^2}B_{B_q}f_{B_q}^2M_{B_q}m_t^2|\vtb^*V_{tq}|^2\eta_B
        \frac{S(x_t)}{x_t},
\label{eq:delmq}
\ee
where $x_t=(m_t/M_W)^2$ and $S(x_t)$ is the Inami-Lim function that results
from evaluation of the box diagrams with internal top quarks~\cite{bb:mix-inamilim}.
A good working approximation~\cite{bb:mix-buras-3} to $S(x_t)$ is
\be
S(x_t) = 0.784\,x_t^{0.76}.
\label{eq:sapprox}
\ee
Buras \etal~\cite{bb:mix-buras-2} have evaluated the QCD correction factor
$\eta_B=0.55$, which is the same for $\bd$ and $\bs$ 
mixing~\cite{bb:mix-buras-3}.  They also stress that the running top mass
$\bar{m}_t(m_t)$, not the pole mass, should be used in the evaluation of
(\ref{eq:delmq}). From the precise determination~\cite{bb:pauls-talk} of 
$m_t=175\pm6$ by CDF and D0, Buras obtains~\cite{bb:andrzejs-talk}
$\bar{m}_t(m_t)=167\pm6$ GeV.

With the top quark mass now so well known, the largest uncertainties in the
evaluation of (\ref{eq:delmq}) arise from the dearth of precise knowledge
about $f_B$, the $B$ decay constant, and $B_B$, the nonperturbative correction
factor.  Flynn~\cite{bb:jonathons-talk} has summarized the lattice
calculations of both quantities for the $\bd$ system.  I will use Buras' 
evaluation~\cite{bb:andrzejs-talk-b} of results from lattice, 
QCD sum rule and QCD dispersion relation calculations:
\be
f_{\bd}\sqrt{B_{\bd}} = 200\pm40 \,\,{\rm MeV}.
\label{eq:fbb}
\ee

Many of the calculational uncertainties are reduced if one considers the
ratio
\be
\xi = \frac{f_{\bs}\sqrt{B_{\bs}}}{f_{\bd}\sqrt{B_{\bd}}}.
\label{eq:xidef}
\ee
Thus the ratio
\be
\frac{\delms}{\delmd} = \xi^2\frac{M_{\bs}}{M_{\bd}}
\left|\frac{\vts}{\vtd}\right|^2
\label{eq:delm-ratio}
\ee
provides a powerful way to obtain $|\vts/\vtd|$.  Both quenched lattice
calculations, summarized here by Flynn~\cite{bb:jonathons-talk}, and QCD sum
rule calculations~\cite{bb:narison-mix-1,bb:narison-mix-2} are consistent with
$\xi=1.15\pm0.05$.  With this precision, we shall see that even the current 
limit on $\delms$ begins to provide interesting constraints in
Unitary Triangle analyses.  One word of caution, however.  Flynn does note
that the unquenching of lattice calculations is likely to increase $\xi$,
perhaps by as much as 10\%. Values as large as 1.25-1.30 can not be
discounted.

\begin{figure}[tb]
\centering
\psfig{figure=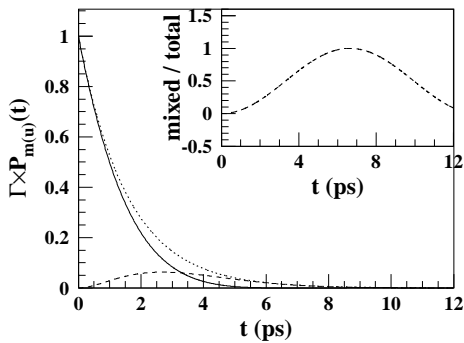,width=6.0cm}
\caption{Probability distributions $\prob_u(t)$ (solid) and $\prob_m(t)$ 
(dashed) for an initially pure $\bod$ or $\bbd$ state and
for a state that does not mix (dotted).  Inset: the fraction
of decays from the mixed state as a function of proper time.}
\label{fig:bdmix}
\end{figure}

This review focuses on measurements that are sensitive to the time
dependence of $\prob_m(t)$, and hence extract $\delm$ directly from the
observed time-dependence.  Figure~\ref{fig:bdmix} illustrates the behavior of the $\bd$
system; the time-dependent measurements examine variables proportional to the
fraction of mixed events as a function of proper time.
Because $x_d=\delmd/\gamd\sim0.75$, measurements are sensitive
to roughly the end of the first $1/2$ cycle (about 6.5 ps), where the
decay rate has decreased by a factor of 65.

For the $\bs$ system, the lower limits on $\delms$
imply that $\chi_s$ will be very close to the
limit $0.5$.  Hence, accurate determination of $x_s$ from measurement of
the total mixed fraction $\chi_s$ is not practical.  A large $\delms$ means a
rapid oscillation frequency --- for $\delms\sim10\,\hbar\,{\rm ps}^{-1}$,
the half cycle would be about $0.3{\rm\,ps}$.  This is about the
current experimental time resolution, making determination of
$\delms$ a challenge on all fronts!

\subsubsection{$B_d-\bar{B}_d$ mixing}

The LEP experiments have done an extraordinary job of determining $\delmd$.
The techniques will be described briefly, with note to novel features of new
analyses from SLAC and CDF. Wu~\cite{bb:mix-wu} has reviewed the LEP measurement 
techniques in greater detail than possible here.

For a time-dependent mixing measurement one must measure
the $B$ proper time $t=LM_B/p_B$.  The large boosts of the $B$ mesons and the
silicon vertex detectors at the LEP experiments, at SLD and at CDF make the
measurement of the decay length $L$ with the requisite precision possible.

Generally, the decay length $L$ is reconstructed by (i) reconstructing a
tertiary charm vertex, either inclusively (eg., with a topological algorithm)
or via an exclusive $D$ channel, (ii) extrapolating the reconstructed $D$ back
to form a vertex with a track, such as a high $p_t$ lepton, or tracks that are
candidate $B$ daughters, and (iii) comparing this secondary vertex to the
primary event vertex.  Depending on the experiment and the type of silicon
detector, either the vertexing is done in the transverse plane and corrected
to a 3 dimensional distance, or via true three dimensional vertexing. 
The resolution on the decay length $L$ in the central
core of the distributions (about 50\% of the area) range from about 90 microns
at CDF and 170 microns at SLD to 250-400 microns at LEP.  Tails on the 
distributions can be parameterized as a Gaussian with widths ranging from 500 
microns to about 1 mm.

Most $\delmd$ analyses
convert the $B$ decay length $L$ back to a proper time using an estimate of the
$B$ momentum.  The sophistication of the estimate ranges from simply taking a
fixed fraction of the beam energy~\cite{bb:mix-delphi,bb:mix-l3} 
to using the sum of all the $B$ decay products, including estimates of the
neutral energy contributions and, for semileptonic modes, an
estimate  of the neutrino momentum via missing energy
constraints~\cite{bb:mix-delphi,bb:mix-aleph,bb:mix-opal-dilepton}.  Resolutions
range from 10\% to 20\%.

With the decay length and $B$ momentum measurements combined, the proper
time resolutions vary from 0.2 ps to 0.3 ps, better than 
20\% of the $\bd$ or $\bs$ lifetime.

To determine the mixing fraction we must tag the flavor of the $B$ meson
at the time of production (the ``production tag'') and at the time of
its decay (the ``decay tag'').  The purity of the $\bo$ content and
of the production and decay flavor tags is quite important. For example, the
statistics
of the samples for the different combinations of 
decay-tag-method/production-tag-method vary from a high of about 60,000
events for the most inclusive method ($\ell/\qj$, explained below) to
about 5000 events ($\ell/\ell$) to a low of several hundred events
($D+\ell/\qj$).  The most inclusive, high statistics samples also have the
lowest $\bo$ purities and highest mistag probabilities, which dilute their
statistical power. The final $\delmd$ sensitivities of the different
tagging techniques are remarkably comparable in the end.

\begin{figure}[tb]
\centering
\psfig{figure=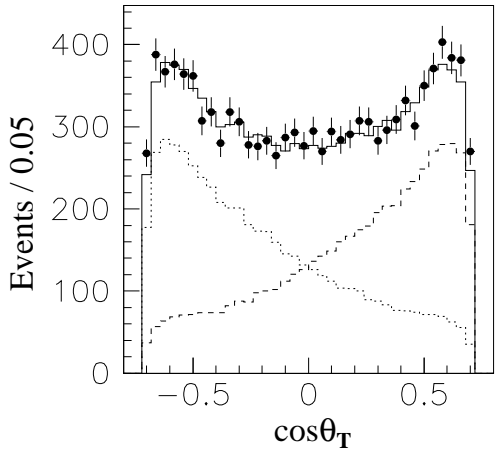,width=6.54cm}
\caption{The $\cos\theta_T$ distribution observed at SLD (points), with
the Monte Carlo prediction of the distributions for $b$ quarks (dashed),
$\bar{b}$ quarks (dotted) and their sum (solid) overlayed.}
\label{fig:afb}
\end{figure}

The new measurements by the SLD
collaboration~\cite{bb:sld-mix-lept,bb:sld-mix-topo,bb:sld-mix-dipo} provide
another example of the power of clean tagging and $b$ purity.  While their
sample of $Z^0$ decays is 20 times smaller than the individual LEP
experiments, the $\delmd$ sensitivity of an individual measurement is within a
factor of two of individual LEP measurements.  SLD takes advantage of
the polarized beams at SLC, which produce a forward-backward asymmetry
\be
\tilde{A}_{FB}=2A_b\frac{A_e+P_e}{1+A_eP_e}
                   \frac{cos\theta_T}{1+\cos^2\theta_T},
\ee
in $\zo\to b\bar{b}$ decays, where $A_e=0.155$ and $A_b=0.94$ specify the
parity violation at the $Ze^+e^-$ and $Zb\bar{b}$ vertices, respectively,
$P_e$ is the beam polarization, and $\theta_T$ is the angle between the thrust
axis and the electron beam direction.  The observed thrust angle distribution,
with the expected asymmetric $b$ and $\bar{b}$ distributions
superimposed, is shown in Figure~\ref{fig:afb}.  By combining the
$\cos\theta_T$ information with other tag methods, SLD has increased the
purity of their production flavor tag.

SLD has also boosted its $b$-quark purity to 93\% in samples with 
inclusively reconstructed secondary vertices by
requiring the mass $M$ and the transverse momentum $P_T$ of the vertex to satisfy
\be
M_{P_T} = \sqrt{M^2+P_T^2}+\left|P_T\right| > 2\,\,{\rm GeV}.
\ee
The decaying hadron's mass must be at least $M_{P_T}$ to produce a vertex
with the observed mass and $P_T$.  They have also increased the $\bo$
purity of their samples by examining the total vertex charge.  Use of 
these techniques has made these first SLD measurements competitive, with
more data still to come.

\begin{figure}[tb]
\centering
\psfig{figure=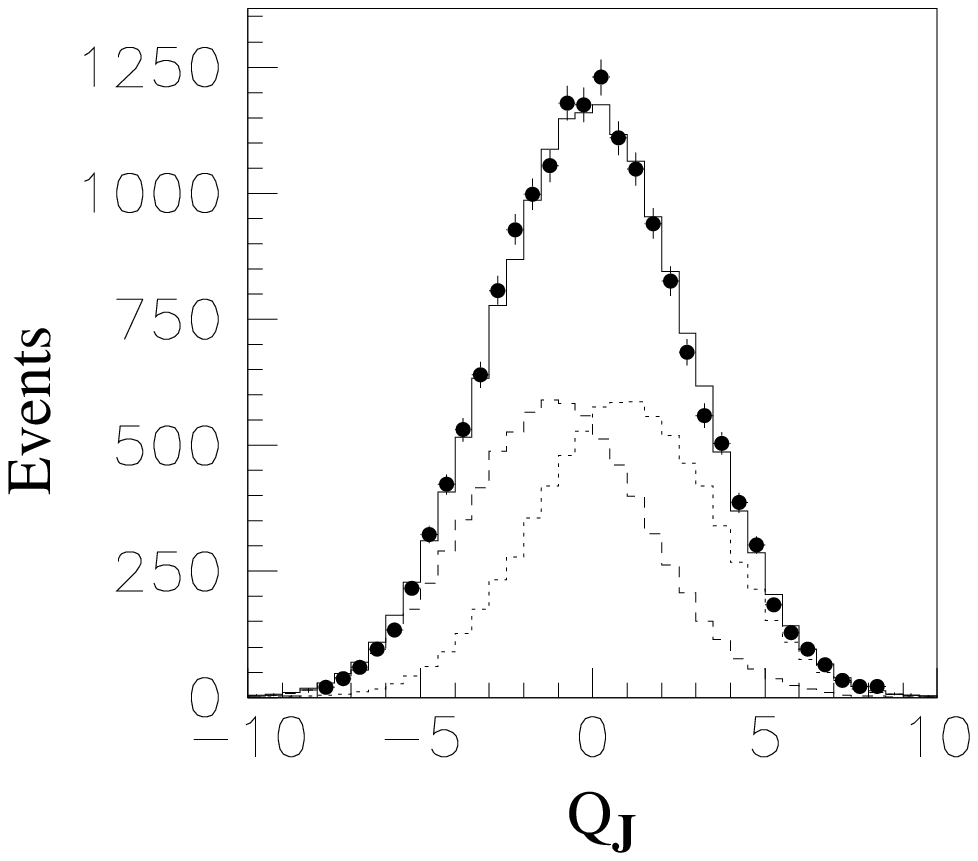,width=7.0cm}
\caption{The jet charge $\qjo$ distribution observed at SLD (points), with
the Monte Carlo prediction of the distributions for $b$ quarks (dashed),
$\bar{b}$ quarks (dotted) and their sum (solid) overlayed.}
\label{fig:jet-charge}
\end{figure}

Most production flavor tags rely primarily on the fact that
the quarks are produced in $b\bar{b}$ pairs, and tag with variables sensitive
to the $b$ quark flavor in the hemisphere (or jet) opposite that of the
decaying $B$ candidate.
Other production tags are based on fragmentation particles in
the same hemisphere as the decaying $B$,
which retain some information about the initial $b$ quark charge.
A summary of the production tags follows:
\begin{list}{}{\setlength{\parsep}{0.3\parskip}\setlength{\leftmargin}{0pt}%
\setlength{\labelwidth}{0pt}\setlength{\itemindent}{1.5em}%
\setlength{\itemsep}{0pt}}
\item[{\bf\boldmath $\qj$ -- jet charge:}] The jet charge is a weighted sum
over the charges of the $n_J$ tracks within a hemisphere: $\qj\sim
\sum_{i=1}^{n_J} q_i |\vec{p}_i\cdot\hat{e}|^\kappa$, where $q_i$ and
$\vec{p}_i$ are the charge and momentum of track $i$, and $\hat{e}$ is an
estimate of the jet direction.  For the jet charge $\qjo$ of the opposite
hemisphere, the weight $\kappa$ is chosen near $0.5$ for ALEPH, DELPHI and
SLD, and at 1 for OPAL.  This tends to weight the charge of $B$ decay products
most heavily (wanted when there is a charged $B$
in the opposite hemisphere), while still retaining the information content of
the fragmentation particles (needed when the opposite $B$ is neutral). The SLD
$\qjo$ distribution is shown in Figure~\ref{fig:jet-charge}.  
For the jet charge $\qjs$ of the hemisphere containing the decaying $B$, 
experiments take $\kappa=0$.
Neutral $B$'s contribute zero net charge, so the sum reduces to a
sum of the fragmentation particles and the production tag is not biased by the
decay flavor.  Some analyses use
$\qj=\qjo-\alpha\qjs$, with a weight $\alpha$ optimized with Monte Carlo.  Jet
charge tagging purities are on the order of 65\% to 70\%.
\item[{\bf\boldmath $\ell$ -- high $p_t$ lepton:}]  A lepton in the opposite
jet with large $p_t$ relative to the jet axis tags the charge of the
$b$ quark.  Such leptons preferentially select $B$ mesons, enhancing
sample purity.
\item[{\bf\boldmath $\pi_{ss}$ -- same-side tagging}]  A candidate
leading fragmentation particle is chosen based on momentum and direction
to the $B$.  For $\bo$ candidates, a $\pi^+$ is
favored as the leading fragmentation particle, since the $\bar{b}$ quark
``pairs'' with the $d$-quark from a $d\bar{d}$ pair, with the $\bar{d}$
left to form a $\pi^+$.  The same sign correlation is expected from $B^{**}$
decays. CDF~\cite{bb:cdf-mix-ss} finds an excess of 
correct-sign tags over wrong-sign tags of about 22\% in their sample.
For $\bs$ mixing, the sign of the leading charged kaon in the fragmentation
products tags the production flavor.
\item[{\bf\boldmath Pol -- polarization:}]  As discussed above, SLD can use 
their forward backward asymmetry to enhance
tagging purity.  The measurement is independent of the $\qj$ measurement,
and they combine the two to obtain a production tagging purity of
84\%.
\end{list}
Some analyses use a combination of several tags to determine the production
flavor.

The $b$ quark flavor at the time of decay is determined by examining
the charge of one or more of the decay products.  The various decay tags
include
\begin{list}{}{\setlength{\parsep}{0.3\parskip}\setlength{\leftmargin}{0pt}%
\setlength{\labelwidth}{0pt}\setlength{\itemindent}{1.5em}%
\setlength{\itemsep}{0pt}}
\item[{\boldmath\bf $\ell$ -- high $p_t$ lepton:}] As above.
\item[{\boldmath\bf $D^{*\pm}$:}]  $D^{*\pm}$ mesons are fully reconstructed
and their charge used to tag the decay flavor.
\item[{\boldmath\bf $\pi^*+\ell$:}] $D^{*\pm}X\ell^-\nu$ are partially
reconstructed using the lepton and the slow charged pion ($\pi^*$) from the 
$D^{*\pm}$ decay.
\item[{\boldmath\bf $K$:}]  The sign of charged kaons from the decay
chain $b\to c\to s$ tags the decay flavor.
\item[{\boldmath\bf $\delta q$ - charge dipole moment:}]  For neutral $B$
decays involving a $D^\pm$, the $D^\pm$ lifetime induces a charge separation
between the $B$ decay vertex and the $D^\pm$ vertex.  The sign of the dipole
moment evaluated along $B$ flight direction tags the production flavor.
\end{list}
Again, the decay tag sometimes consists of a combination of several methods.

\begin{figure}[tb]
\centering
\psfig{figure=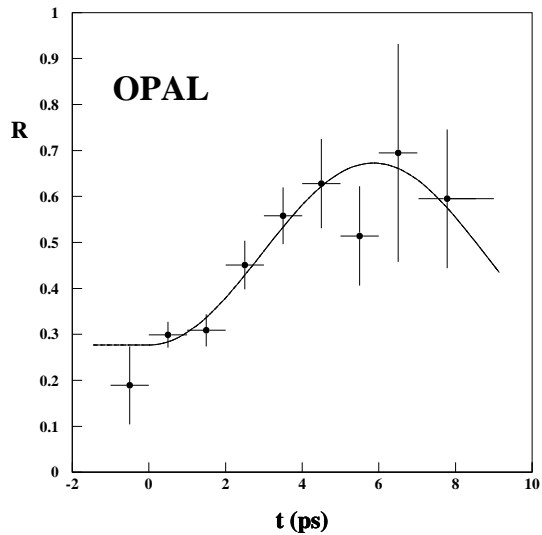,width=7.0cm}
\caption{The mixed event fraction $R$ as a function of proper time for
the OPAL $D^{*\pm}+\ell/\qj$ analysis.  The curve is the result of the
fit.}
\label{fig:opalmix}
\end{figure}

\begin{figure}[tb]
\centering
\psfig{figure=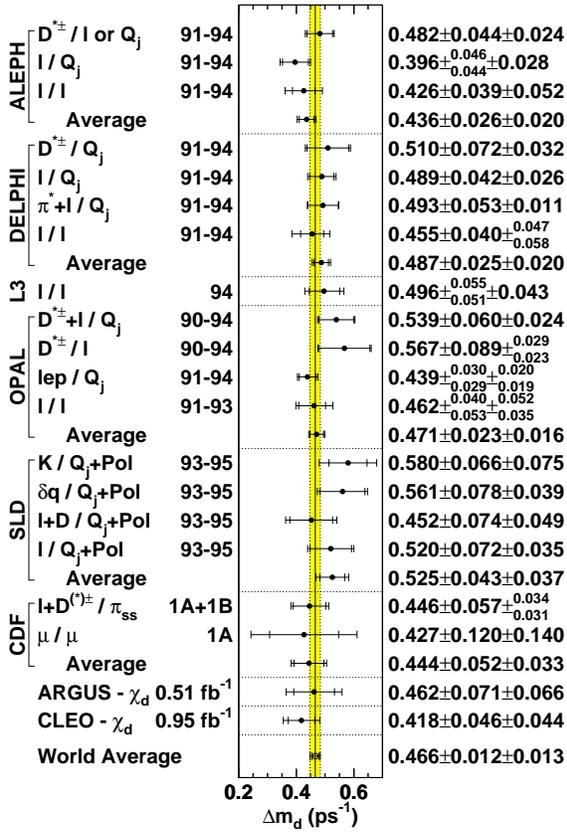,width=7.4cm}
\caption{Summary of $\delmd$ measurements.  The grey band indicates the
world average. Method and data 
sample are listed at left, the results at right.}
\label{fig:md_avg}
\end{figure}

Based on the production and decay flavor tags, the fraction of mixed events
as a function of proper time is determined.  
The resulting distribution for the OPAL $D^{*\pm}+\ell/\qj$ 
analysis~\cite{bb:mix-opal-dstlep} is shown in Figure~\ref{fig:opalmix}.
The first half cycle, with the statistical precision degraded by the
end, is clearly visible.  These distributions are fit to a signal (mixing)
term governed by $\prob_m(t)$ (Eq.~\ref{eq:mixprob}) convoluted with a
resolution function, plus background terms for non-$\bo$ contributions and
mistagging.  The $\delmd$ results from the time dependent measurements of
ALEPH~\cite{bb:mix-aleph}, DELPHI~\cite{bb:mix-delphi}, 
L3~\cite{bb:mix-l3}, 
OPAL~\cite{bb:mix-opal-dilepton,bb:mix-opal-dstlep,bb:mix-opal-lQ},
SLD~\cite{bb:sld-mix-lept,bb:sld-mix-topo,bb:sld-mix-dipo} and
CDF~\cite{bb:cdf-mix-ss,bb:cdf-mix-dilep} at the time of the conference
are summarized in Figure~\ref{fig:md_avg}.
The results based on measurements
of $\chi_d$ at the $\Upsilon(4S)$~\cite{bb:mix-cleo,bb:argus-summary}
and the averages for each experiment are also shown.
The world average is $\delmd=0.466\pm0.012\pm0.013\,\hbar\,{\rm ps}^{-1}$.
The precision is under 4\%.

In the averaging, correlated systematic errors must be addressed.
Ignoring all correlations is unrealistic; assuming
that similar techniques have fully correlated systematics will
probably overestimate the correlations, and will miss correlations that exist
between the other measurements.  
For each technique, only a subset of the errors categorized by all experiments
has been considered in the individual measurements. To compensate
for this somewhat, I have made coarse categories of correlated categories:
\newcounter{syst}
\begin{list}%
{\arabic{syst}.\ }{\usecounter{syst}%
\setlength{\parsep}{0.3\parskip}\setlength{\leftmargin}{0pt}%
\setlength{\labelwidth}{0pt}\setlength{\itemindent}{1.5em}%
\setlength{\itemsep}{0pt}}
\item Lifetimes of $B^0$, $B^+$, $B_s$, and $b$ baryons. Contributions 
from different lifetimes
within a $\delmd$ measurement are added in quadrature and this total error 
is taken as the correlated error when comparing measurements.
\item Similarly for $D$ lifetimes.
\item The fraction of $b$ baryons produced in $b\bar{b}$ hadronization.
\item The fraction of $B_s$ produced in $b\bar{b}$ hadronization.
\item The $c\bar{c}$ fraction of $\zo$ decays.
\item Similarly for the total fraction of $usd$.
\item The average $B$ momentum at the $\zo$, {\it ie.}, 
$<p_B>/E_{\rm \scriptsize beam}\sim 70\%$ at LEP.
\item For leptons from $b$, the fraction $f_{bc}$ from the cascade
$b\to c\to l$.
\item In the cascade $b\to c\to l$, the fraction from $\bd$ and from charged
$B$'s.
\item  Uncertainties from the fraction of $D^{**}$'s in $B$ decays, and
for analyses that reconstruct $D^{*}$'s, the fraction contributed by
$\bp$.
\item $\delms$.
\item Uncertainty in the charged and neutral $B$ production fraction
in the $\chi_d$ measurements at the $\Upsilon(4S)$.
\end{list}
The total systematic for each measurement in each category is shown
in Table~\ref{tab:mdsyst}.  If an experiment has already combined some or all
of their measurements, I have listed the systematic for that combined 
measurement.

Were the correlations ignored, the systematic error would have dropped to
$0.009\,\hbar\,{\rm ps}^{-1}$.  Even though the correlated errors
considered are relatively minor, they increase the error by 50\%!.  A LEP
group has formed to perform a more systematic averaging of $\delmd$.
These results indicate that for individual measurements,
a systematic uncertainty can be neglected only if small on the scale of the 
statistical error of that world average, not if it is small only on the 
scale of the uncertainty in that particular measurement.

\begin{table*}[tb]
\caption{The correlated systematic on $\delmd$ in each of the 12 categories
for the different experiments (or measurements).  The total systematic is
shown for comparison. The units are $\ten{-3}\,\hbar\,{\rm ps}^{-1}$.}
\label{tab:mdsyst}
\centering
\setlength{\tabcolsep}{1mm}
\begin{tabular}{cc|c|rrrrrrrrrrrr}\hline\hline
Experiment & Method & Total & \multicolumn{12}{c}{Systematic in Correlated Category} \\ \cline{4-15}
 &      & systematic & 1 & 2 & 3 & 4 & 5 & 6 & 7 & 8 & 9 & 10 & 11 & 12 \\ 
\hline
ALEPH  & $D^{\pm}$/$\ell$ or $\qj$ 
&   24&   3&   2&   0&   0&   0&   0&   7&   0&   0&  18&   0&   0 \\
ALEPH & $\ell$/$\qj$,$\ell$/$\ell$
&   30&  16&   0&   5&   5&   0&   0&   0&  21&   3&   0&   4&   0 \\
DELPHI & Comb.
&   20&   6&   0&   5&   9&   1&   0&   4&  10&   3&   6&   0&   0 \\
L3 & $\ell$/$\ell$
&   43&   4&   0&  17&  37&   1&   1&  15&   0&   0&   0&   1&   0 \\
OPAL & $D^{\pm}$+$\ell$/$\qj$ 
&   24&   7&   0&   0&   0&   0&   0&   0&   0&   0&  19&   0&   0 \\
OPAL & $D^{\pm}$/$\ell$
&   ${}^{+29}_{-23}$
      &   7&   0&   0&   8&  14&   0&   0&   0&   0&   9&   0&   0 \\
OPAL & $\ell$/$\qj$
&   19&   5&   0&   5&   4&   0&   0&   0&   2&   0&   0&   0&   0 \\
OPAL & $\ell$/$\ell$)
&   ${}^{+52}_{-35}$
      &  22&   0&  12&  28&   0&   0&   0&   0&  11&   8&   6&   0 \\
SLD & Comb.
&   37&   9&   7&  10&  10&   0&   0&  10&   5&   3&   2&   4&   0 \\
CDF & $D^{\pm}$+$\ell$/$\pi_{ss}$
&   ${}^{+34}_{-31}$
      &   3&   0&   0&   0&   0&   0&   0&   0&   0&  29&   0&   0 \\
CDF & $\mu/\mu$
&  140&   0&   0&   0&   0&   0&   0&   0&   0& 113&   0&   0&   0 \\
ARGUS & $\chi_d$
&   66&  15&   0&   0&   0&   0&   0&   0&   0&   0&   0&   0&  20 \\
CLEO &  $\chi_d$
&   44&  13&   0&   0&   0&   0&   0&   0&   0&   0&   0&   0&  20 \\ 
\hline\hline
\end{tabular}
\end{table*}

\subsubsection{$B_s-\bar{B}_s$ mixing}
Over the past year there have been two significant developments in the effort
to determine $\delms$.  The oscillations have not yet been observed, but the
lower limit on $\delms$ has improved significantly.

First, the analyses themselves have improved.  The older analysis techniques
were based on the same $\ell/\ell$ and $\ell/\qj$ samples used in the $\delmd$
analyses.  While these inclusive samples have high statistics, the $B_s$
content was low (10\%) and the $\delms$ bounds were sensitive to the
fraction $\fbs$ of $b$ quarks that hadronize to form a $\bs$.  The $\bs$ 
samples have been enriched by using
reconstructed $D_s$ or $\phi$ mesons in conjunction with a high $p_t$
lepton (or high momentum hadron) to tag the decay
flavor.  These analyses have a $\bs$
content ranging from 25\% up to as high as 60\%.  In addition, because of
the exclusively reconstructed final state, the proper time resolution has
improved, which is crucial for observing the rapid oscillation of the 
$\bs$ system.

The second breakthrough is a new technique for extracting the lower
$\delms$ limit.  Previously, the limits were determined from the
likelihood difference curves obtained from fitting to the mixed event
fraction: $-\Delta\log\like(\delms)=\log\like(\delms^{\rm\scriptsize
max}) -\log\like(\delms)$, where $\delms^{\rm\scriptsize max}$ is the
mass difference that maximizes the likelihood. If the oscillation
period is small relative to the proper time resolution, the value of
$\delms^{\rm\scriptsize max}$ tends to be biased~\cite{bb:ms-moser}
towards a frequency at which the true amplitude is comparable to the
statistical noise, so the likelihood function must be ``calibrated''
to determine the correct 95\% C.L. limit.  This has been done for each
measurement by studying many toy experiments generated from a fast MC
to determine the likelihood change that encompasses 95\% of the
experiments at a given $\delms$.

\begin{figure}[tb]
\centering
\psfig{figure=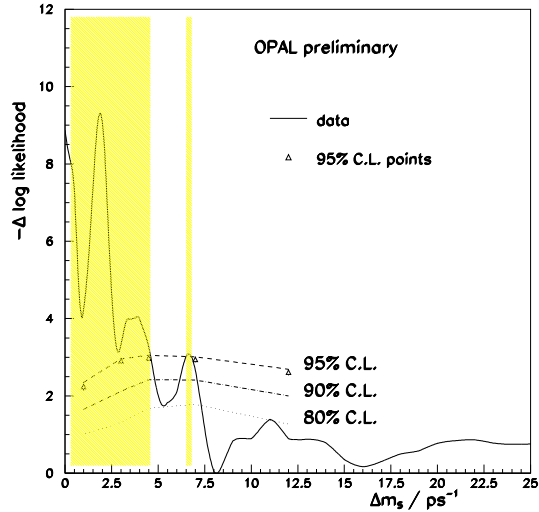,width=7.0cm}
\caption{The OPAL combined likelihood curve (solid) as a function of $\delms$
for their three $\delms$ analyses, and the Monte Carlo calibration curves for
the confidence levels 95\%,90\% and 80\%.  The shaded regions indicate
$\delms$ values excluded at 95\% C.L..}
\label{fig:opalms}
\end{figure}

Combining the $-\Delta\log\like$ curves from different measurements is
difficult.  The resolutions involved are non-Gaussian; the systematics are
correlated between measurements; the curve from each measurement has local
minima and maxima.  Under these conditions, there is not a well-defined
procedure for combining the different curves.
Furthermore, the combined likelihood itself must be recalibrated before a
lower limit can be obtained, which offers a severe impediment to the
combination of results from different experiments.  The OPAL collaboration
has, however, done a beautiful job of combining~\cite{bb:ms-opal-2} their three
$\delms$ likelihood
curves~\cite{bb:mix-opal-dstlep,bb:mix-opal-lQ,bb:ms-opal-2}.  The resulting
likelihood and MC calibration are shown in Figure~\ref{fig:opalms}.

An alternative technique for extracting $\delms$ has been
developed~\cite{bb:ms-moser,bb:ms-aleph-kaon} for ALEPH by Moser and 
Roussarie.  Rather than
extracting a likelihood as a function of $\delms$, their procedure, the
``amplitude method'', is essentially a Fourier-transform.  In
their fit function, they replace the mixed decay probability $\prob_m$ of
Eq.~\ref{eq:mixprob} with
\be
\prob_{m}(t) = \frac{\Gamma}{2}e^{-\Gamma t}[1 - A\cos(\nu t)].
\label{eq:modprob}
\ee 
The frequency $\nu$ is fixed, and one fits for the amplitude $A$ of that
frequency component in the data.  For frequencies $\nu$ far from the true
value of $\delms$, the amplitude should be zero.  As $\nu$ approaches
$\delms$, $A\to 1$.  The distribution of $A$ should be the Fourier-transform
of (\ref{eq:mixprob}), a Breit-Wigner with a width of 
$2/\tau_{\bs}\approx 1.3\,{\rm ps}^{-1}$.  One scans over values of $\nu$,
fitting for the frequency components $A(\nu)$.  $\delms$ is excluded from
frequency ranges over which the amplitudes are less than one at the 95\% C.L.,
{\it ie.} where $A(\nu)+1.645\sigma_A(\nu)<1$.  Both ALEPH~\cite{bb:ms-aleph-3}
and DELPHI~\cite{bb:ms-delphi} have evaluated their $\delms$ limits using this
technique.

\begin{figure}[tb]
\centering
\psfig{figure=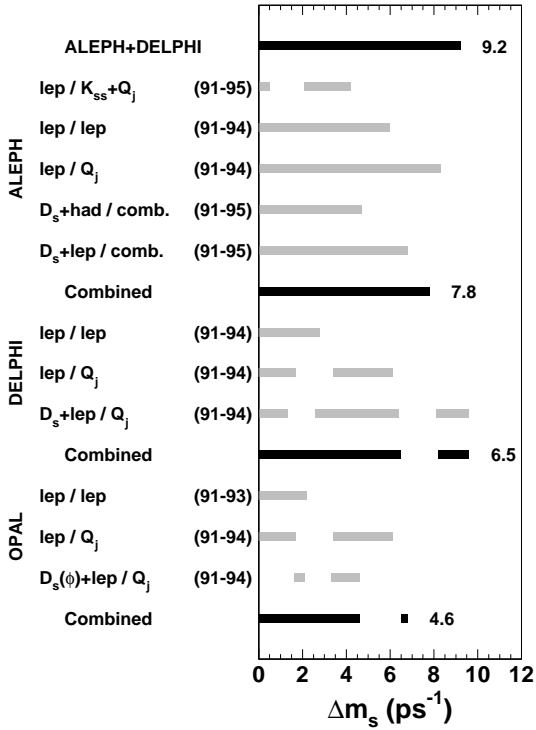,width=7.0cm}
\caption{The $\delms$ regions excluded at the 95\% C.L. by the 
ALEPH, DELPHI and OPAL measurements (grey bands).  Also shown are the
excluded regions from the combined measurements (black bands) of 
ALEPH+DELPHI (amplitude method -- a.m.), 
ALEPH (a.m.), DELPHI (a.m.) and OPAL 
($-\Delta\log\like$), with the 95\% lower $\delms$ limit in each case.}
\label{fig:delmsmeas}
\end{figure}

\begin{figure}[tb]
\centering
\psfig{figure=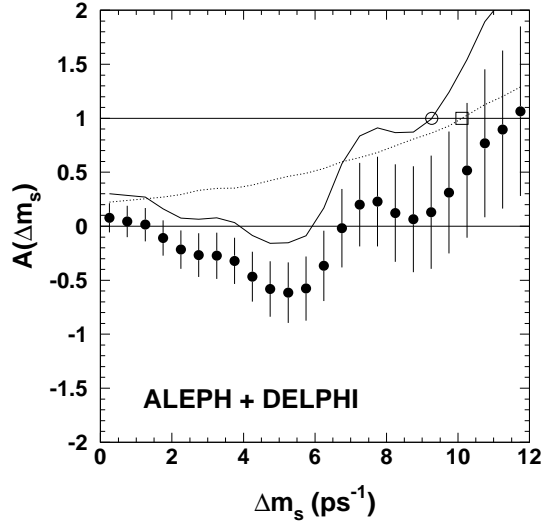,width=7.0cm}
\caption{$A(\delms)$ versus $\delms$ for the combined ALEPH and DELPHI
measurements (solid circles).  The solid curve represents $A+1.645\sigma_A$;
its intersection with 1 (open circle) determines the 95\% C.L. lower bound.
The dotted curve shows $1.645\sigma_A$; its intersection
with 1 (open square) indicates the $\delms$ sensitivity of the combined 
analysis}
\label{fig:ult-ms}
\end{figure}

The results of the ALEPH, DELPHI and OPAL 
$\delms$ analyses are summarized in Figure~\ref{fig:delmsmeas}. The
amplitudes extracted from different measurements are simple to combine ---
$A(\nu)$ at a given frequency is a physical
measurement with a well-defined uncertainty.
ALEPH and DELPHI have
combined all of their measurements in this fashion to obtain improved
lower limits.  Combining $A(\nu)$ from both experiments yields the data shown 
in  Figure~\ref{fig:ult-ms}, from which the lower bound
$\delms>9.2\,\hbar\,{\rm ps}^{-1}$ is obtained.  The amplitude error
provides an estimate of the sensitivity to $\delms$.
For the combined results the estimated sensitivity is 
$10.1\,\hbar\,{\rm ps}^{-1}$, larger than the lower bound.

\subsection{Summary of $|\vtd|$ and $|\vts|$ }

\begin{figure}[tb]
\centering
\psfig{figure=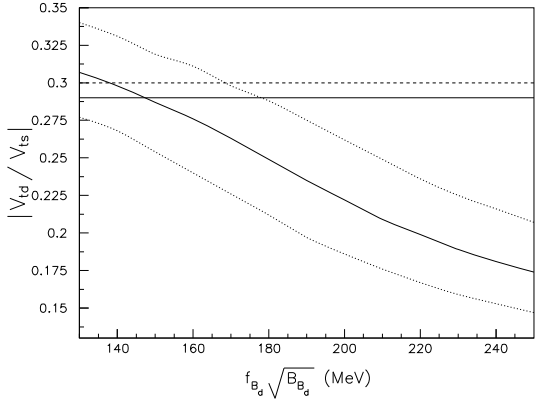,width=7.0cm}
\caption{The one standard deviation bounds (dotted curves)
for $|\vtd/\vts|$ versus $f_{\bd}\sqrt{B_{\bd}}$ from a recent
UT-analysis of Ali and London~\protect\cite{bb:alilon}, assuming $\vts=\vcb$,
which is valid within the Wolfenstein approximation.  The
solid horizontal line is the upper limit on $|\vtd/\vts|$ from the $\delmd$
and $\delms$ results presented here combined with the 95\% C.L. upper limit
on $\xi=1.15\pm0.05$.  The dashed line corresponds $\xi=1.3$.}
\label{fig:alilon}
\end{figure}

While we now have a precise determination of the mass difference
$\delmd=0.466\pm0.018\,\hbar\,{\rm ps}^{-1}$,
uncertainty in $f_{\bd}\sqrt{B_{\bd}}$ currently limits the precision 
in the value of $|\vtd|$ that can be extracted from this measurement.
Combining this result with (\ref{eq:delmq}) and (\ref{eq:fbb}),
one obtains a range for $|\vtd|$ of 0.005 to 0.015.  However, the
95\% C.L. limit $\delms>9.2\,\hbar\,{\rm ps}^{-1}$, combined with this
measurement of $\delmd$, begins to prove interesting.  If we also take
a 95\% C.L. upper limit on $\xi$ ({\em c.f.} (\ref{eq:xidef})) based on
$\xi=1.15\pm0.05$, then Eq.~\ref{eq:delm-ratio} implies $|\vtd/\vts|<0.29$.
This begins to restrict the allowed range of $f_{\bd}\sqrt{B_{\bd}}$ in UT
analyses, as Figure~\ref{fig:alilon} suggests.  Even with $\xi$ as high as
1.3, a 95\% C.L. limit of $|\vtd/\vts|<0.30$ is still constraining.  SLD and
CDF expect to limit $\delms$ with similar sensitivity.

The limits on the rare decay $\kp\to\pnn$ constrain $\vtd$ at the ${\cal
O}(0.1)$ level only.  Data being taken now will significantly improve this
limit.  Comparison of rare $B$ processes $B\to\rhogam/B\to\kstgam$
yields a limit on $|\vtd/\vts|$ in the range $0.45-0.6$, depending upon the
assumptions made and models used in the extraction of the limit.  This is not
far from the limit obtained from $\delms$ studies, and there is more data
available at CLEO for measuring this ratio.  In the long run, both methods
should provide valuable information for $\vtd$, unitary triangle analyses, and
the quest for new physics.

\section{Determining $|\vcb|$}
Improving the precision on $|V_{cb}|$ remains a very important goal.  $|\vcb|$
determines the Wolfenstein parameter $A$ in UT analyses.  Since $A$ appears
raised to large powers in some of the theoretical constraints, the effect of
any uncertainty is magnified.  Burchat and Richman~\cite{bb:jeffs-review}
point out that even though $|V_{cb}|$ is already known quite precisely, its
contribution to the uncertainty in the constraints on $\rho$ and $\eta$ from
the $CP$-violating parameter $\vep$ in $K_L\to\pi\pi$ decay, relative to
the constraints from the $B$ system, rivals the
contribution from the nonperturbative correction $B_K$.

There are two complementary approaches to the precise determination of
$|\vcb|$: inclusive measurement of the total semileptonic $b\to c$
branching fraction, and measurement of the $B\to D^*\ell\nu$ hadronic
form factor $\form$ at the zero recoil point.  Since some of the key inputs
that are used to control the theoretical uncertainties remain to be 
verified experimentally, agreement between the two methods provides a
powerful consistency check.

\subsection{Inclusive measurements}
Extraction of $|\vcb|$ through the inclusive measurement of the total $B$ 
(or $b$) semileptonic branching fraction has the advantage of great 
statistical power.  The semileptonic
partial width can be written as
\be
\Gamma = \Gamma_0\left\{1 - a_1(\frac{m_b^2}{m_c^2})\frac{\alpha_s}{\pi} + 
\frac{a_{\rm np}}{m_b^2} 
+ \orde(\alpha_s^2,\frac{\alpha_s}{m_b^2},\frac{1}{m_b^3})\right\},
\ee
where the $\Gamma_0$ is the tree level rate
\be
\Gamma_0 = \frac{G_F^2m_b^5}{192\pi^3}|\vcb|^2z_0(\frac{m_b^2}{m_c^2}),
\ee
$a_1$ is the first order perturbative QCD correction, $a_{\rm np}$ is
the nonperturbative correction, and $z_0$ is a phase space factor.
The exact form for these terms can be found 
elsewhere~\cite{bb:incb-bigi,bb:incb-bu,bb:incb-ls}. The nonperturbative
correction is small and known to about 10\% of itself, contributing little
to the overall uncertainty in the predicted rate.  One might expect
uncertainty in the $b$ quark mass to introduce large uncertainties in
the total rate via the $m_b^5$ dependence. Recent calculations
by Shifman \etal~\cite{bb:incb-sub} and Ball~\etal~\cite{bb:incb-ball}
of $\Gamma/|\vcb|^2=41.3\pm4\,{\rm ps}^{-1}$ and 
$43.2\pm4.2\,{\rm ps}^{-1}$, respectively, have rather small 
(10\%) uncertainties, though.  Two factors appear to mollify the $b$ mass 
uncertainty. As Uraltsev~\cite{bb:uraltsev-review} discusses,
the variation of $a_1$ with $m_b$
tends to cancel the variation of the tree-level rate $\Gamma_0$.  A significant
reduction also comes~\cite{bb:incb-sub} from constraining the mass 
difference $m_b-m_c$ to the heavy quark expansion prediction
\be
m_b-m_c=\bar{M}_B-\bar{M}_D + \frac{\mu_\pi^2}{2}(\frac{1}{m_c}-\frac{1}{m_b})
 + \cdots,
\ee
where the ``spin-averaged mass'' $\bar{M}_H=\frac14(M_H+3M^*_H)$ and
$\mu_\pi^2$ parameterizes the $b$ quark kinetic energy within the hadron.
An agressive uncertainty, on the order of 50 MeV, was adopted for $m_b$ by 
these authors.
The validity of these constraints have yet to be tested experimentally.
For this review, I will average the two calculations and use
the 10\% uncertainty estimate, $\Gamma/|\vcb|^2 = (42.3\pm4.2)\,{\rm ps}^{-1}$.

\begin{figure}[tb]
\centering
\psfig{figure=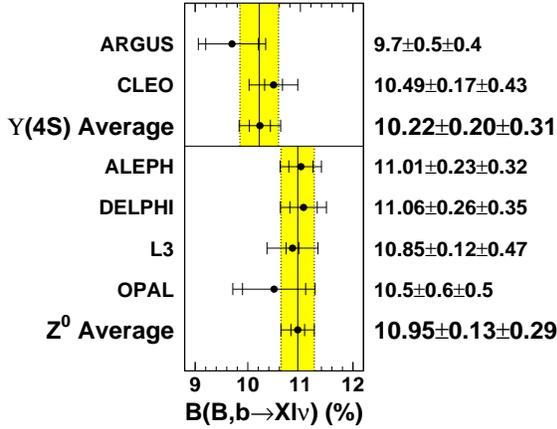,width=7.3cm}
\caption{The inclusive branching ratios $\calb(B\to X\ell\nu)$ and
$\calb(b\to X\ell\nu)$ measured at the $\Upsilon(4S)$ and the $Z^0$,
respectively.  The grey bands indicate the averages.}
\label{fig:inclus}
\end{figure}

Experimentally, the discrepancies between semileptonic rates for the 
$B$ meson, measured~\cite{bb:argus-inc,bb:argus-summary,bb:cleo-inc}
at the $\Upsilon(4S)$, and the $b$ quark,
measured~\cite{bb:aleph-inc,bb:delphi-inc,bb:l3-inc,bb:opal-inc}
at the $Z^0$, have decreased.  For the $\Upsilon(4S)$ the dilepton 
measurements are quoted, as preferred by those experiments, since 
while statistical uncertainties are larger than in the single
lepton measurements, the model dependence is much smaller.  I quote
the lifetime-tagged ALEPH result for the same reason.
Figure~\ref{fig:inclus} summarizes the current situation.  The lifetime-tagged
ALEPH analysis and the new L3 results in particular have helped to
alleviate the discrepancy.

Taken at face value, the branching ratios agree within 1.5 standard
deviations.  However, given the short $b$ baryon lifetimes, one would
actually expect the LEP value to be {\em lower} than that measured at
the $\Upsilon(4S)$, and the real discrepancy is more severe.

It should be noted that these measurements are the combination of the 2
processes $b\to c\ell\nu$ and $b\to u\ell\nu$.  Rosner
estimates~\cite{bb:inc-rosner} the $b\to u$ fraction to be $(1.85{\rm\ to\
}2.44)|\vub/\vcb|^2$, or about $(1.5\pm1.0)$\%.  While this seems small, {\em
it is of the same order as the experimental statistical uncertainties} and
therefore should be corrected for.  After correction, the average $B$ meson and
$b$ quark charmed semileptonic branching fractions are $(10.07\pm0.38)$\% and
$(10.79\pm0.33)$\%, respectively. The statistical, systematic and $|\vub|$
correction uncertainties have been combined in quadrature.

These measurements, the average
lifetimes~\cite{bb:pdg96} at the $\Upsilon(4S)$ and $Z^0$ of
$<\!\tau_B\!>\,=1.59\pm0.04$ ps and $<\tau_b>=1.55\pm0.02$ ps, and the rate
prediction combine to give
\begin{center}
\begin{tabular}{l@{ : $|\vcb|=\,$}l}
$\Upsilon(4S)$ & $(38.7\pm0.9\pm1.9)\tten{-3}$ \\
$\zo$          & $(40.6\pm0.7\pm2.0)\tten{-3}$ \\
\end{tabular}
\end{center}

\subsection{$B\to D^*\ell\nu$}
Measurement of the $B\to D^*\ell\nu$ differential decay rate at the zero
recoil point currently provides the determination of $|\vcb|$ with the
least theoretical uncertainty.  The experimental techniques have been
reviewed in some 
detail~\cite{bb:ritchie-glasgow,bb:sheldon-dpf,bb:tomasz-leppho,bb:wagner}.  
The measurements are reviewed briefly here,
with the focus on an improved averaging procedure that accounts for
the correlation between the extracted form factor slopes and $|\vcb|$.

The differential decay rate for $B\to D^*\ell\nu$ is given by
\be
\frac{d\Gamma}{dw}=\frac{G_F^2}{48\pi^3}\kappa(m_B,m_D,w)|\vcb|^2\form^2(w),
\label{eq:excl-rate}
\ee
where $w = v_B\cdot v_{\ds} = (m_B^2+m_{\ds}^2-q^2)/2m_Bm_{\ds}$, 
and is the boost of the $D^*$ in the $B$ rest frame, $q^2$ is the
four-momentum transfer to the leptonic system, and $\kappa$ is a known
function.  In general, the form factor $\form(w)$
cannot be calculated in a model-independent fashion.  

At the zero-recoil point $w=1$, however, heavy quark effective theory
(HQET) allows calculation of $\form(1)$ with controllable theoretical
uncertainties.  HQET and experimental tests are discussed in more
detail in the talks of Richman~\cite{bb:jeffs-talk} and
Martinelli~\cite{bb:guidos-talk} and, for example, in various reviews
by Neubert~\cite{bb:neubert-review,bb:neubert-lecture}.  HQET predicts
that $\form(1)\to 1$ for transitions
between heavy quarks, with corrections at the level of $1/m_q$ or smaller.  
At this zero recoil point, the heavy quark changes flavor 
without perturbing its color field, so there is little suppression of the
decay rate at that point.

Experimentally, then, one must determine the form factor at $w=1$.  
Unfortunately, the differential decay rate
rate vanishes at $w=1$, so $\form(w)$ must be measured nearby and
extrapolated to $w=1$.  Since the true function $\form(w)$ is unknown,
it is expanded around the zero recoil point,
\be
\form(w)=\form(1)[1-\rhohat^2(w-1)+\chat(w-1)^2+\cdots].
\ee
While experiments have attempted to extract both the slope and the
curvature, they are not yet sufficiently sensitive to
constrain $\chat$, and therefore quote results for strictly linear fits.
A positive curvature must exist, but only a small bias is introduced
from the linear restriction
because the physically allowed range for $w$, from 1 to 1.5, is rather
small. The bias will be estimated later.

The decay $B\to\ds\ell\nu$ is favored over $D\ell\nu$ for several reasons.
First of all, as Luke~\cite{bb:lukes-theorem} noted, the 
$\orde(\Lambda_{\rm\scriptsize QCD}/m_b)$ corrections to $\form(1)$ vanish
for $\ds\ell\nu$.  The calculations for $\form(1)$ for
$B\to\ds\ell\nu$ have also been much more thoroughly scrutinized.
Experimentally, the $\ds\ell\nu$ mode has the largest branching fraction,
and it is not helicity-suppressed near $w=1$, while $D\ell\nu$ is.
Hence a more precise and reliable extrapolation to $w=1$ is possible for 
$\ds\ell\nu$.

\begin{figure}[tb]
\centering
\psfig{figure=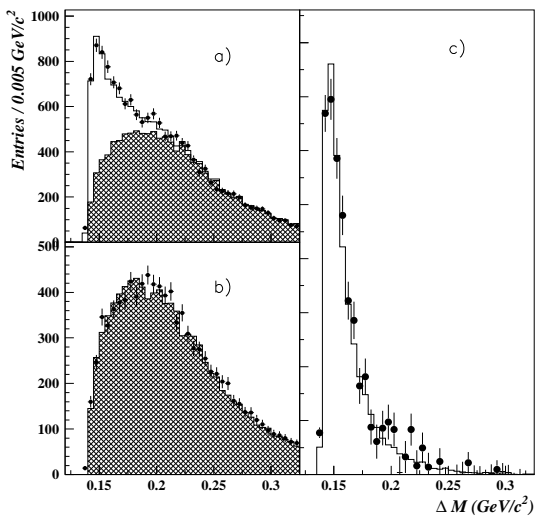,width=7.0cm}
\caption{Reconstructed mass difference for DELPHI inclusive $D^0$ $|\vcb|$
analysis. (a) Right sign $\ell^\pm\pi^\mp$ combinations in data (points),
background MC (dashed histogram) and signal MC (open histogram). (b)
Wrong sign $\ell^\pm\pi^\pm$ combinations. (c) Data after simulated background
subtraction.}
\label{fig:delphi-mass}
\end{figure}

Most
measurements~\cite{bb:argus-summary,bb:aleph-excl,bb:cleo-excl,bb:delphi-excl,bb:opal-excl}
have used the decays $\bo\to D^{*-}\ell^+\nu$, $D^{*-}\to\pi^-\bar{D}^0$,
explicitly reconstructing $\bar{D}^0\to K^+\pi^-(n\pi)$.  
DELPHI~\cite{bb:delphi-excl}
has, in additional, inclusively reconstructed the $D^0$ meson, to obtain one of
the most precise recent measurements of $\form(1)|\vcb|$.  While their $D^0$
mass resolution of 700 MeV is quite broad, they do see a clear $\ds$ signal in
the mass difference distribution $\Delta M=M_{D^0\pi}-M_{D^0}$
(Figure~\ref{fig:delphi-mass}).  CLEO~\cite{bb:cleo-excl} has used $\bm\to
D^{*0}\ell^-\bar{\nu}$, $D^{*0}\to D^0\pi^0$ in addition to the neutral $B$
decays.

With an undetected neutrino in the final state, estimating $w$ is
challenging.  At the $\Upsilon(4S)$, the $B$ mesons are nearly at rest, so
$w=E_{\ds}/M_{\ds}$ is an excellent approximation, yielding a
$w$ resolution of about 4\% of the total 1-1.5 range.  At the $\zo$, the
$B$ energy is not known, but with the large boosts, $w$ can be estimated 
from the $B$ flight direction and missing energy
constraints.  The resulting $w$ resolutions at LEP are about 15\% to 20\% 
of the total range for the exclusive analyses, and about 20\% to 40\% for the
DELPHI inclusive analysis.  The reconstructed $w$ distributions are then fit 
using (\ref{eq:excl-rate}) to obtain $\form(1)|\vcb|^2$ and $\rhohat$,
properly accounting for the smearing of $w$.

The $\Upsilon(4S)$ and $\zo$ measurements complement each other nicely.
Because of the very
soft $B$'s at the $\Upsilon(4S)$, the $w$ resolution is excellent and simple
lepton momentum and kinematic requirements reduce $D^{**}$ backgrounds to about
5\%.  On the other hand, the low $B$ momentum means that the charged $\pi$'s
from the $D^*$ decays are very soft near $w=1$, resulting in poor
reconstruction efficiency where the information is most important for the
extrapolation to $w=1$.
Because of the large boost of the $B$'s at LEP, the
$D^*$ reconstruction efficiency is good near $w=1$.  The $w$ resolution,
however, is much worse, and the $D^{**}$ backgrounds are severe -- of
order 15\%.  This is unfortunate, since we are now only beginning to
obtain reliable information concerning $B$ decays to $D^{**}$.

\begin{table*}[tb]
\caption{Results from fits to the observed $w$ distributions for 
$B\to D^*\ell\nu$.  The parameters $\rho_{\rm \scriptsize stat}$ and
$\rho_{\rm \scriptsize sys}$ are the statistical and systematic
correlation coefficients between $F(1)|\vcb|/\ten{-3}$ and $\rhohat$. All
intercepts have been scaled to a common set of branching ratios and
lifetimes (see text). For the DELPHI-avg, the ``statistical' correlation
coefficient listed is the total correlation coefficient for the averaged
exclusive and inclusive results, excluding the small (0.04) common systematic.
The results for $B\to D\ell\nu$ are also listed.}
\label{tab:dstfits}
\centering
\begin{tabular}{llrrrr} \hline\hline
& & $F(1)|\vcb|/\ten{-3}$ & $\rhohat$ & $\rho_{\rm \scriptsize stat}$ &
$\rho_{\rm \scriptsize sys}$  \\ \hline
$D^*\ell\nu$
&ALEPH       & $31.0\pm1.8\pm2.0$ & $0.29\pm0.18\pm0.12$ & 0.921 & 0.612 \\
&DELPHI-incl & $36.4\pm2.2\pm2.6$ & $0.74\pm0.20\pm^{0.18}_{0.16}$ & 0.96 & 0.816 \\
&DELPHI-excl & $35.2\pm3.5\pm3.0$ & $0.77\pm0.26\pm0.10$ & 0.906 & 0.616 \\
&DELPHI-avg  & $36.5\pm2.1\pm2.2$ & $0.78\pm0.16\pm0.10\pm0.04$ & 0.869 & 0.0 \\
&OPAL        & $33.9\pm2.9\pm2.3$ & $0.44\pm0.24\pm0.12$ & 0.95 & 0.577 \\
&ARGUS       & $39.0\pm3.9\pm2.8$ & $1.17\pm0.24\pm0.06$ & 0.90 & 0.601 \\
&CLEO        & $35.1\pm1.9\pm1.8$ & $0.84\pm0.12\pm0.08$ & 0.90 & 0.610 \\ 
\hline
$D\ell\nu$ 
&CLEO - $\nu$ & $36.2\pm5.7\pm4.2$ & $0.73\pm0.24\pm0.10$ & 0.940 & 0.725 \\
&CLEO - kinem.& $32.7\pm6.0\pm5.3$ & $0.52\pm0.29\pm0.10$ & 0.972 & 0.574 \\
&ALEPH        & $38.8\pm6.7\pm6.9$ & $0.00\pm0.49\pm0.38$ & 0.986 & 0.953 \\
\hline\hline
\end{tabular}
\end{table*}

The experimental results are summarized in Table~\ref{tab:dstfits}.  All
results have been updated to the following common set of constants:
\begin{center}
\begin{tabular}{l@{ : }l}
$R_b$~\cite{bb:aleph-excl} & $(22.09\pm0.21)$\% \\
$\calb(b\to B^0)$~\cite{bb:pdg96} & $(37.8\pm2.2)$\% \\
$\tau_{\bo}$~\cite{bb:kroll} & $(1.56\pm0.05)$ ps \\
$\tau_{\bp}/\tau_{\bo}$~\cite{bb:kroll} & $1.02\pm0.04$ \\
$\calb(D^0\to K^-\pi^+)$~\cite{bb:jeffs-talk} & $(3.88\pm0.1)$\% \\
\end{tabular}
\end{center}
The statistical correlation coefficients from the 
individual fits are quite large. (Note: ARGUS has not
published its coefficients -- I have assumed that they are similar to CLEO's. 
OPAL did not list the systematic uncertainties on their slope, so 
$\rho_{\rm \scriptsize sys}$ is estimated to be similar to the other LEP
coefficients.) If statistical fluctuations or systematic biases have resulted
in a low slope in the fit, the intercept will be lowered as well.  Hence the
intercepts should not be averaged without regard to the slopes.

\begin{figure}[tb]
\centering
\psfig{figure=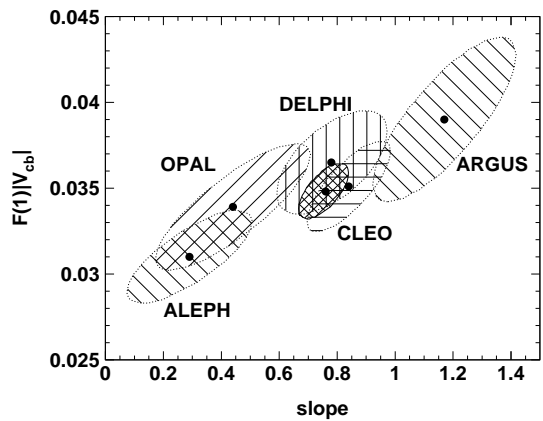,width=7.0cm}
\caption{One standard deviation error ellipses for $\form(1)|\vcb|$ versus
the form factor slope $\rhohat^2$.  The black cross-hatched ellipse is the
average.}
\label{fig:dst-ellipse}
\end{figure}

The one standard deviation error ellipses are shown for the five
experiments in Figure~\ref{fig:dst-ellipse}. While the intercepts
appear consistent, the measured slopes do not --- they are consistent
only at the 5\% confidence level.  Note that a precise measurement of
the form factors by CLEO~\cite{bb:cleo-ff}, for which backgrounds have
been highly suppressed, implies that the average slope measured in these
analyses should be $\rhohat\approx 0.77\pm0.22\pm0.07$.
An averaging of the slopes and the intercepts simulataneously with
their correlations and also correlated systematic errors between
experiments taken into account yields
$\form(1)|\vcb|=0.0348\pm0.0016$. The $\chi^2$ is 12.2 for 8
degrees of freedom.  Since the errors in $|\vcb|$ should reflect the
unsettling state of the slopes, I've chosen to adopt the Particle Data
Group procedure and to scale the error by $\sqrt{12.2/8}$.  The final
result is then $\form(1)|\vcb|=0.0348\pm0.0020$ The average slope,
also with errors scaled, is $\rhohat=0.75\pm0.11$.

We must address the bias induced by the linear form factor parameterization
in the fits.  Patterson~\cite{bb:ritchie-glasgow} and
Stone~\cite{bb:sheldon-dpf} have evaluated the possible shifts in a variety of
models, and find that corrections of up to $+4\%$ might be necessary.
ALEPH~\cite{bb:aleph-excl} has fit using constraints on the
curvature $\chat$ versus $\rhohat$ estimated~\cite{bb:cabrini}by Cabrini
and Neubert.  While their central value does not change, the statistical
uncertainty increases by 4\%.  It seems that an additional uncertainty of 4\%
is appropriate for the linear bias.  For an estimate of the bias, I used
the Cabrini--Neubert relationship to find the curvature that would yield
on average the slope observed in a linear fit.  This procedure implies an
average bias in the fits of -2.5\%.  After a $(2.5\pm^4_{2.5})$\% correction,
$\form(1)|\vcb|=(35.7\pm2.1\pm^{1.4}_{0.9})\tten{-3}$, where the errors are 
the experimental uncertainty and the bias uncertainty,
respectively.

To extract $|\vcb|$, we need a prediction for $F(1)=\eta_A(1+\delta_{1/m^2})$,
where $\eta_A$ is the perturbative QCD correction, and $\delta_{1/m^2}$
incorporates the $1/m_b^2$ and $1/m_c^2$ corrections.  Recently,
Czarnecki~\cite{bb:czar} has completed a complete two-loop calculation of the
perturbative correction, finding $\eta_A=0.960\pm0.007$.  Neubert has
summarized~\cite{bb:neubert-review} the work done by various 
authors~\cite{bb:incb-sub,bb:falkneub,bb:mannel,bb:neubert} for the
$1/m^2$ correction, finding $\delta_{1/m^2}=(-5.5\pm2.5)$\%.  The predictions 
of HQET are now just beginning to be tested,  for example
through the measurement of the individual form factors in $B\to D^*\ell\nu$
decays~\cite{bb:cleo-ff}.

The total correction is therefore $\form(1)=0.91\pm0.03$, from which we
obtain
\be
|\vcb|=0.0391\pm0.0027\pm0.0013,
\ee
where the experimental and bias 
uncertainties have been combined in the first error, and the second error is
the uncertainty in $\form(1)$.

\subsection{$B\to D\ell\nu$}
Both CLEO~\cite{bb:cleo-dlnu} and ALEPH~\cite{bb:aleph-excl} have
results for similar form factor studies of the decay $B^0\to D^-\ell^+\nu$.
CLEO has performed two analyses that have similar sensitivities and largely
independent samples, one based on ``neutrino-reconstruction'' techniques
that will be discussed below, and one based on the $D$--$\ell$ kinematics
assuming the $B$ mesons produced at the $\Upsilon(4S)$ are at rest.
The ALEPH analysis is similar to its $D^*\ell\nu$ analysis.

\begin{figure}[tb]
\centering
\psfig{figure=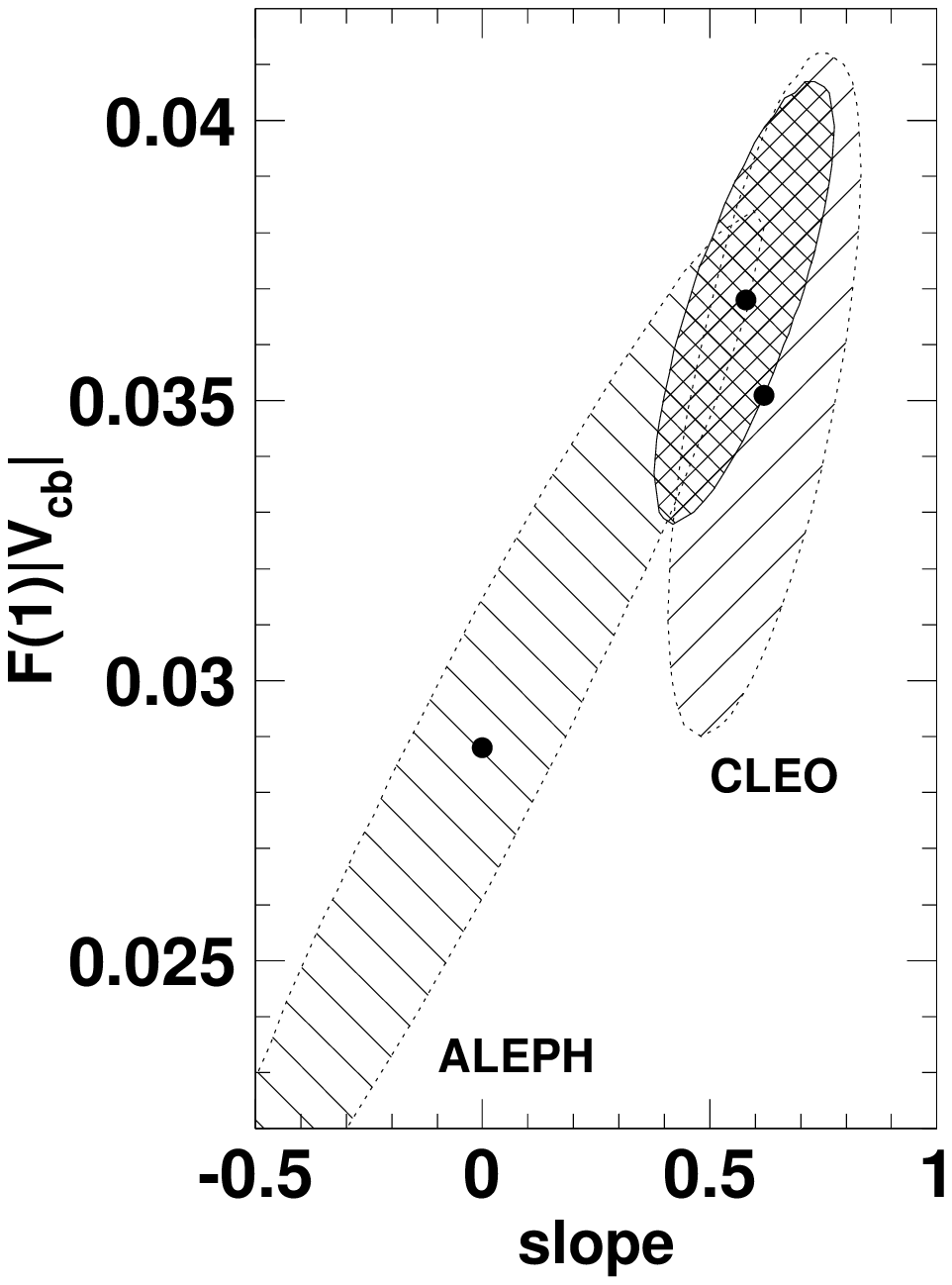,width=7.0cm}
\caption{One standard deviation error ellipses for $\form(1)|\vcb|$ versus
the form factor slope $\rhohat^2$ for $B^0\to D^-\ell^+\nu$.  The black 
cross-hatched ellipse is the average.}
\label{fig:dlnu-ellipse}
\end{figure}

The results, after correcting to the standard branching ratios listed above,
are also summarized in Table~\ref{tab:dstfits}.  The error ellipses for
both experiments are shown in Figure~\ref{fig:dlnu-ellipse}. 
The CLEO results have been reaveraged with the
slope and intercept correlations taken into account along with the
common systematic uncertainties.
The simultaneous average of the three
slope and intercept measurements gives 
$\form(1)|\vcb|=(36.8\pm4.0)\tten{-3}$ and
$\rhohat=0.58\pm0.20$.  For the intercept in particular, it is clear from 
Figure~\ref{fig:dlnu-ellipse} that a naive average would have resulted in a 
very biased
result --- almost a full standard deviation lower than the results obtained
here.

While the prediction of $\form(1)$ for $D\ell\nu$ has not received nearly the
scrutiny as the $D^*\ell\nu$ prediction, a recent calculation
predicts~\cite{bb:cabrini,bb:nir} $\form(1)=0.98\pm0.07$.  Taking this
calculation, and assuming the same bias from the linear form factor
parameterization as in the $D^{*}\ell\nu$ case, we have
$|\vcb|=0.0385\pm0.0045\pm0.0028$.

\subsection{Summary}
\begin{figure}[tb]
\centering
\psfig{figure=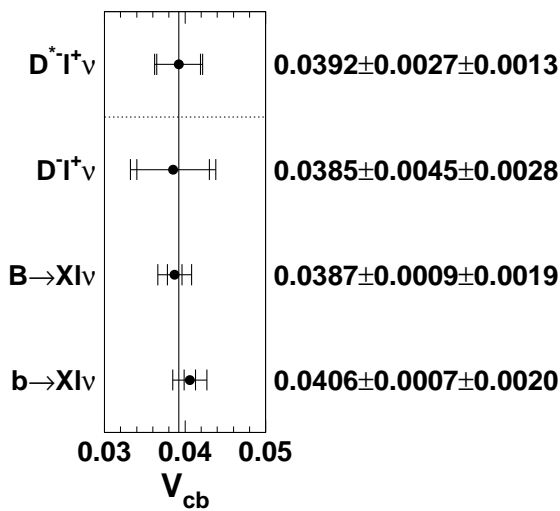,width=7.3cm}
\caption{Summary of the $|\vcb|$ measurements.}
\label{fig:vcbsumm}
\end{figure}

The $\vcb$ values obtained from inclusive measurements at the $\Upsilon(4S)$
and at LEP and from the exclusive $B\to D^{(*)}\ell\nu$ decays are shown in
Figure~\ref{fig:vcbsumm}.  The level of agreement between the different
inclusive and exclusive measurements is quite good.  The theoretical
uncertainties in the inclusive and exclusive measurements are quite different,
so the agreement is certainly heartening, and suggests that, at least at the
5\% level, the theoretical calculations are in reasonable control.  However,
given the broad range of estimates for the theoretical
uncertainties~\cite{bb:uraltsev-review,bb:neubert-review}, and the infancy of
the experimental checks of the theoretical inputs, it seems premature to
average the results.

\section{Determining $V_{ub}$}
Until recently, the sole evidence for $b\to u$ transitions has come from the
observation~\cite{bb:CLEO-btou-a,bb:ARGUS-btou} of leptons with momenta
in a range accessible to $b\to u\ell\nu$, yet high enough
to be reached only rarely in the dominant $b\to c\ell\nu$ process.
These studies have yielded the oft-quoted value $|\vub/\vcb|=0.08\pm0.02$.
Model dependence dominates the uncertainty --- the details of
hadronization significantly affect this endpoint region, and the rate into
this range cannot be calculated reliably.  

Experimentally, the rate measurements themselves show significant variation
within a single model.
Patterson~\cite{bb:ritchies-talk} has re-evaluated those analyses: the
discrepancies appear to arise from the procedure used to extract
$|\vub/\vcb|$ from the
observed $b\to u\ell\nu$ rate. This was accomplished using the measured rate
for inclusive $b\to c\ell\nu$ decays with leptons in a moderately high
momentum range, allowing $|\vub/\vcb|$ to be obtained directly.  The procedure
relied on modeling of both $b\to u\ell\nu$ and $b\to c\ell\nu$.  If instead
the $B$ lifetime, now known quite precisely, is used to directly extract
$|\vub|$, the experimental discrepancies vanish.  The new average is
compatible with the standard $|\vub/\vcb|$ value in use,
and the resolution of this problem lends considerable confidence that the the
experimental uncertainty is reasonable.

ALEPH has presented~\cite{bb:kroha} first evidence for observation of
semileptonic $b\to u$ transitions at LEP.  The large $b\to c$ background
must be understood at the 1\% level for the reliable extraction of
$|\vub|$, which seems difficult given our incomplete  knowledge of
the charm semileptonic decays of $B$ hadrons.

Exclusive charmless semileptonic decays provide an alternate route to
$|\vub|$.  The theoretical road is still rocky, since the form factors
for the exclusive decays must be known both for the
determination of detection efficiencies and for the rate predictions
needed to extract $|\vub|$.  Because these are heavy$\to$light
transitions, HQET cannot help as it did for $b\to c\ell\nu$, and we must
live with model dependence for now. There is great
activity in the calculation of form factors spanning a variety of models and 
techniques, and consistency among the different technique will help to lend
some confidence in the $|\vub|$ extraction. 

\begin{figure}[tb]
\centering
\psfig{figure=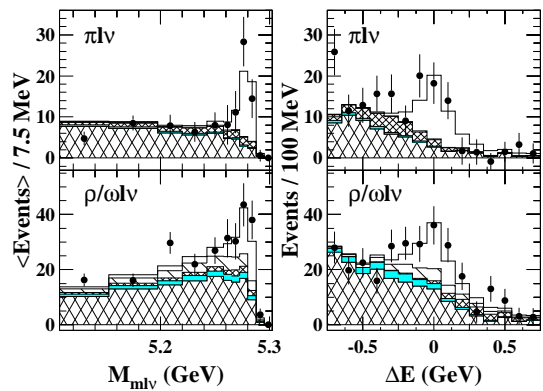,width=7.0cm}
\caption{$\mcand$ (left) and $\dele$ (right) distributions in
the $\Delta E$ and $M_{m\ell\nu}$ signal bands,
respectively.
The data (points) have 
continuum and fake-lepton backgrounds subtracted. The coarse 
crosshatch, grey and unshaded components are $b \to cX$ background, 
feed-down background from other $B\to X_u\ell\nu$ decays, and signal,
respectively. For the $\pi$ ($\rho/\omega$) modes, the fine crosshatch 
shows the
signal mode cross-feed level
$\rho/\omega\to\pi$ ($\pi\to\rho/\omega$) and the single hatch,
$\pi\to\pi$ ($\rho/\omega\to\rho/\omega$) cross-feed.}
\label{fig:pirho-sig}
\end{figure}

CLEO~\cite{bb:ritchies-talk,bb:cleo-pirho} has recently extracted $|\vub|$ 
from the study of the $B\to\pln$ and $B\to\rwln$ decays of both $\bp$ and 
$\bo$.  To suppress the
large backgrounds from the dominant $b\to c\ell\nu$ processes, they use the
missing energy $\emiss$ and momentum $\pmiss$ in the entire event to infer
information about the missing neutrino.   Selection requirements --- zero
observed charge, identically one observed lepton, 
$\mmiss\equiv\emiss^2-\pmiss^2\sim m_\nu^2=0$ --- suppress events in which
$\pmiss$ would misrepresent a true primary neutrino.  They find resolutions
on the neutrino momentum and direction of 110 MeV (roughly 5\%) and $6\deg$,
respectively.  They can then fully reconstruct signal candidates, examining
$\mcand\equiv(\ebeam^2-|\vec{p}_m+\vec{p}_\ell+\vec{p}_\nu|^2)^{1/2}$,
which should peak at the $B$ mass, and $\dele\equiv\ecand-\ebeam$, which
should peak at zero.  They see clear evidence for a signal in both
variables (Figure~\ref{fig:pirho-sig}) in both the $\pi$ modes and the
$\rho/\omega$ modes.  CLEO extracts branching ratios and $|\vub|$ from
fits to data in the region $|\Delta E|<0.75$ GeV and 
$5.1075\le M_{m\ell\nu}<5.2875$ GeV. 

CLEO has evaluated their experimental efficiencies using several 
models~\cite{bb:models-1} that span a variety of calculation techniques.  For the
branching ratios, the model dependence is moderate because
the efficiencies do not depend on the overall normalization of the form
factors.  CLEO can test the validity of a given model by comparing the 
measured ratio of branching fractions
$\calb(B\to\rln)/\calb(B\to\pln)$, obtained with efficiencies determined
from the model, to the ratio predicted by that model.
The K\"{o}rner
and Schuler model was only consistent at the $0.5\%$ level, so it was excluded
from any model averages.  From the remaining models they obtained
$\calb(B^0\to\pi^-\ell^+\nu)=(1.8\pm0.4\pm0.3\pm0.2)\times 10^{-4}$ and
$\calb(B^0\to\rho^-\ell^+\nu)=(2.5\pm0.4^{+0.5}_{-0.7}\pm0.5)\times 10^{-4}$,
where the errors are statistical, systematic and the estimated 
model-dependence, respectively. The asymmetric error in the $\rho$ modes
arises from the uncertainty in  nonresonant $B\to\pi\pi\ell\nu$ background,
which CLEO has limited by studying $\pi^0\pi^0\ell\nu$.

\begin{figure}[tb]
\centering
\psfig{figure=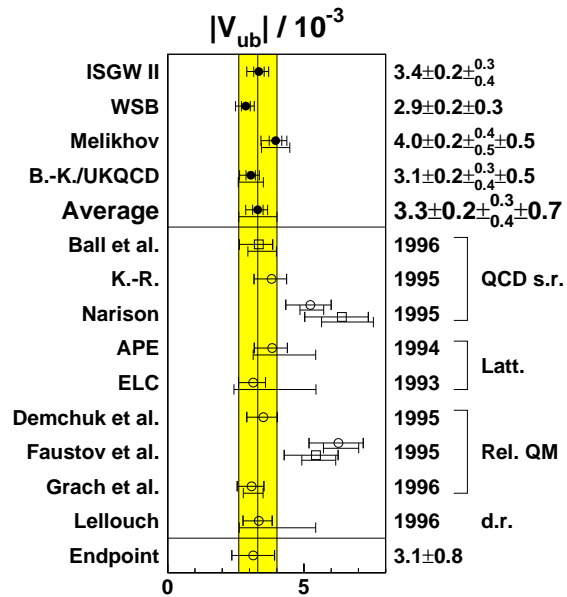,width=7.3cm}
\caption{Values for $|\vub|$ extracted from the exclusive CLEO measurements.
Top section:  Combined $\pln$ and $\rln$ results from models considered 
by CLEO. Middle section:  Results based on CLEO model-averaged branching 
fractions and predictions from a sampling of other calculations based on 
QCD sum rules (QCD s.r.), lattice calculations (Latt.), relativistic
quark models (rel. QM) and dispersion relations (d.r.).  B.--K. refers
to Burdman-Kambor; K.-R. refers to Khodjamirian--R\"{u}ckl.
Open circles (squares)
are $\pln$ ($\rln$) results.  Bottom section: Inclusive endpoint value.
The points show the experimental errors. In the middle section, the
estimated model dependence is included. An error
bar underneath the data point indicates {\em that model's} internal error.
The grey band show CLEO's estimated model dependence.}
\label{fig:vubres}
\end{figure}

Extraction of $|\vub|$ has greater model dependence because the absolute form
factor normalization must be known.  To combine the information from the $\pi$
and $\rho/\omega$ modes within a model, CLEO determines $|\vub|$ by
constraining the $\rho/\pi$ ratio to that predicted by the model.  The
resulting values are shown in the top of Figure~\ref{fig:vubres}, from which
CLEO estimates $|V_{ub}|=(3.3\pm0.2^{+0.3}_{-0.4}\pm0.7)\times 10^{-3}$, where
the errors are statistical, systematic, and estimated model dependence,
respectively.  Comparisons of a larger sampling of recent
models~\cite{bb:models-2}, from which $|\vub|$ has been extracted using the
above branching fractions, are also shown.  They generally agree with CLEO's
value, and suggest that the estimated model dependence is reasonable.

This new determination of $|\vub|$ is in excellent agreement with the value 
obtained from the endpoint analysis, which bolsters confidence in
the value used for the past few years.
At the moment, it is not clear how to properly average these results, since
the errors in both are dominated by theoretical uncertainties, uncertainties
that in this case are correlated.  The recent CLEO result is perhaps
the most robust estimate to date.

\section{CPLEAR -- $CP$ and $CPT$}
This year marks the close of data taking for the CPLEAR experiment.  The
many results of this beautiful experiment are summarized in these Proceedings
by B.~Pagels~\cite{bb:pagels}.  They have made many precise determinations of
the $CP$-violating parameters in the neutral kaon system.

They have also made significant contributions to tests of
$CPT$ conservation.  There had been a long-standing two standard deviation
discrepancy
between the world average of $\phi_{+-}$  (the phase of the $CP$-violating
parameter $\eta_{+-}$ in $K_L\to \pi^+\pi^-$ decay) and the ``superweak
phase'' 
$\phi_{\rm\scriptsize sw}\equiv\tan^{-1}(\frac{2\Delta m}{\Gamma_S-\Gamma_L})$.
Such a discrepancy would signal $CPT$ violation.
Measurements by the FNAL E731 experiment~\cite{bb:e731} indicated that the
problem lay with a high world average of the $K_L-K_S$ mass difference,
$\Delta m$, with which $\phi_{+-}$ measurements are highly correlated.  Among
CPLEAR's measurements are $\Delta m=(529.2\pm1.8\pm1.5)\tten{-7}\,\hbar\,{\rm
s}^{-1}$, the most precise to date, and
$\phi_{+-}=43.5\deg\pm0.5\deg\pm0.5\deg\pm0.4\deg$.  These measurements,
along with other recent results~\cite{bb:roy}, have confirmed this suggestion, and
$CPT$ invariance is alive and well.

CPLEAR also has an indication of a $T$-violating asymmetry in $K$ semileptonic
decay based on a fraction of their data, with the assumption that $CPT$
is conserved in the semileptonic decay amplitudes.  Such an asymmetry is
expected to accompany the known $CP$-violation in the neutral $K$ system.  It
will be exciting to see the results of this analysis from the full data
sample.

\section{Conclusion}
In summary, the year has continued to benefit from the ingenuity brought to
bear on weak quark mixing.
We now know $\delmd$ with outstanding precision.  I look
forward to improved calculations of $f_B$ and $B_B$ from the lattice and 
elsewhere.  Limits on $\delms$ are beginning
to provide useful constraints in unitary triangle analyses, and improved
statistics for $\delms$ and for rare $B$ and $K$ decays hold great promise.

Values of $|\vcb|$ show excellent agreement in results obtained
from inclusive and exclusive studies.  Extraction of $|\vcb|$ from
$D^*\ell\nu$ decays remains the most reliable determination.  This method
is currently statistically limited, and improvements should be seen soon.
With the improving precision, proper handling of correlations is
necessary, and understanding the experimental discrepancies in the form
factor slopes is crucial.  The unknown form factor curvatures could soon
become a limiting factor in the precision.

Studies of $|\vub|$ have entered a new era with the first determination from
exclusive channels.  These have bolstered our confidence in the
inclusive determination, and provide a testing ground for exclusive models.
This will help, in turn, reduce uncertainties in the inclusive determination.

Impressive as the precision of the measurements is becoming,
the Standard Model remains unscathed.  The next few years,
as analyses of current experiments finish and as high luminosity $B$
experiments begin data-taking, hold the promise of much excitement.

\section*{Acknowledgments}
I would like to thank S.~Dong, J.~Kroll, S.~Manly, H.-G.~Moser, P.~Roudeau and
S.~Willocq for their help, discussions and results concerning the 
time-dependent mixing analyses, D.~Rousseau and F.~Simonetto for their
assistance with $|\vcb|$ results, and B.~Pagels and L.~Littenburg for
much useful information concerning their experiments.  I would also
like to thank my fellow rapporteurs, particularly J.~Richman and
A.~Buras, for their help and many interesting discussions.  Finally, I
would like to thank the conference organizers for their hospitality and the
fine organization of a most enjoyable conference.

\end{document}